\documentclass[conference]{IEEEtran}

\usepackage{graphicx}
\usepackage{subfigure}
\usepackage{amsmath}
\usepackage{braket}
\usepackage{algorithm, algorithmicx, algpseudocode}
\usepackage{comment}
\usepackage{tikz}
\usepackage{amssymb} 
\usepackage{booktabs} 
\usepackage{array} 
\newcommand{\tikzmark}[2]{
    \tikz[overlay,remember picture,baseline] 
    \node[anchor=base] (#1) {$#2$};
}
\usepackage{listings}
\usepackage{seqsplit}
\usepackage{hyperref}

\usepackage{pdfpages}

\newcommand{\todo}[1]{}
\renewcommand{\todo}[1]{{\color{red} TODO: {#1}}}

\begin{document}

\title{DiaQ: Efficient State-Vector Quantum Simulation}

\author{
  \IEEEauthorblockN{Srikar Chundury\IEEEauthorrefmark{1}, Jiajia Li\IEEEauthorrefmark{1}, In-Saeng Suh\IEEEauthorrefmark{2}, Frank Mueller\IEEEauthorrefmark{1}}
  \IEEEauthorblockA{\{schundu3, jli256, fmuelle\}@ncsu.edu, suhi@ornl.gov}
  \IEEEauthorblockA{\IEEEauthorrefmark{1}North Carolina State University, Raleigh, NC 27695, USA}
  \IEEEauthorblockA{\IEEEauthorrefmark{2}Oak Ridge National Laboratory, Oak Ridge, TN 37831, USA}
}


\maketitle

\begin{abstract}

  In the current era of Noisy Intermediate Scale Quantum (NISQ)
  computing, efficient digital simulation of quantum systems holds
  significant importance for quantum algorithm development,
  verification and validation. However, analysis of sparsity within
  these simulations remains largely unexplored. In this paper, we
  present a novel observation regarding the prevalent sparsity
  patterns inherent in quantum circuits. We introduce DiaQ, a new
  sparse matrix format tailored to exploit this quantum-specific
  sparsity, thereby enhancing simulation performance. Our contribution
  extends to the development of libdiaq, a numerical library
  implemented in C++ with OpenMP for multi-core acceleration and SIMD
  vectorization, featuring essential mathematical kernels for digital
  quantum simulations. Furthermore, we integrate DiaQ with SV-Sim, a
  state vector simulator, yielding substantial performance
  improvements across various quantum circuits (e.g., $\sim$26.67\%
  for GHZ-28 and $\sim$32.72\% for QFT-29 with multi-core
  parallelization and SIMD vectorization on Frontier). Evaluations
  conducted on benchmarks from SupermarQ and QASMBench
  demonstrate that DiaQ represents a significant step towards
  achieving highly efficient quantum simulations.

\end{abstract}


\section{Introduction}

Quantum Computing has progressed significantly in the past decade with
a variety of algorithms like Variational Quantum Eigen-Solvers (VQE),
Quantum Approximate Optimization Algorithms (QAOA), Quantum Neural
Networks (QNN) in many fields like
cryptography~\cite{Shor_1997,anschuetz2018variational},
optimization~\cite{farhi2014quantum}~\cite{Peruzzo_2014},
physics/chemistry~\cite{Low_2019,Uvarov_2020}, machine
learning~\cite{Biamonte_2017,Harrow_2009}, and
finance~\cite{Woerner_2019,Braine_2021}.

The concept of mimicking quantum behavior (quantum systems and their
dynamics) and using it for ``quantum parallelization'', where multiple
states can co-exist at the same time, is called quantum
simulation. Quantum simulation is a key area of focus of quantum
computing because it paves the path for solving classically infeasible
problems using quantum ``behavior'' (quantum dynamics). This has been
used in fields like material science and machine
learning~\cite{analog_qc_material_science}.

In 1982, Richard Feynman hypothesized about having a universal quantum
simulator~\cite{Feynman1982Simulating}, a device that could simulate
any quantum system. Broadly, they are two distinct types of quantum
simulators: analogue and digital.

\textbf{Analogue Quantum Simulator:}
The process of simulating quantum behavior by physically implementing
quantum dynamics is called analogue quantum simulation. In contrast to
Feynman's vision, analogue simulators today simulate limited quantum
systems. They excel in optimization problems, quantum chemistry
simulations, and machine learning tasks, many-body systems etc., but
cannot be generalized easily. Notwithstanding its restricted
applications, this class of simulation is looked at as the most
feasible with today's quantum hardware.

\textbf{Digital Quantum Simulator:}
Quantum simulation (also referred as quantum computing today) can be
truly universal when looked at as quantum circuits, a sequence of
quantum gates. These ``gates'' are norm-maintaining transformations of
qubits in a quantum system. Digital quantum simulators are quantum
devices that can be ``programmed'' (using these gates) into any other
quantum system.

We are in the Noisy-Intermediate-Scale-Quantum (NISQ) era, where
quantum hardware is still error-prone, with limited coherence times
and costly quantum error correction. There are many programmable
devices on the market today, IBM's superconducting systems, Quera's
Rydberg atoms, Ion-Q's Ion-trap machines, among others.

Today, we make use of classical linear-algebra tools and
techniques for algorithmic verification as NISQ devices remain
noisy. Simulation also serves as a means for device verification. The
tools that are used for this purpose are also called simulators, but
are classical in nature. For these classical simulators of digital
quantum simulators, quantum gates are represented as matrices, and
quantum circuits are a sequence of such gate matrices. From now on,
when we use the term ``simulation'' we refer to the classical
simulation of digital quantum simulation/computing.

Run-time is our primary metric for comparing simulation
techniques. This choice is driven by the exponential increase in
computational demands with quantum circuit width and linear growth
with depth. The efficient simulation speed is crucial for exploring
variational quantum algorithms and directly impacts the cost and
effectiveness of algorithm development. Hence, improvement in one
iteration of such algorithms has a multiplied effect on the overall
solution finding process.

In search for ways to improve today's simulations, this paper makes
the following contributions:
\begin{itemize}
\item Identification of diagonal sparsity in quantum circuit unitaries;
\item formulation of the DiaQ format;
\item implementation of the libdiaq C++ library (with
  python wrappers);
\item integration of libdiaq with SV-Sim~\cite{li2021svsim}; and
\item evaluation of SV-Sim + libdiaq on multi-core CPUs with
  vectorization, with comparisons against the default dense version.
\end{itemize}
\section{Background}

\subsection{Quantum Simulations}

Different types of quantum simulations are distinguished by their
functionality for an $n$-qubit system.

\textbf{State-Vector (SV) Simulation:}
The state of a quantum system is represented as a vector of size
$2^{n}$. This vector stores the amplitudes of all the possible states
of the quantum system. For instance, a 2-qubit system's vector stores
amplitudes of $\ket{00}$, $\ket{01}$, $\ket{11}$ and $\ket{10}$. And
for a 3-qubit system, the state vector stores amplitudes of
$\ket{000}$, $\ket{001}$, $\ket{010}$ and so on. The initial state of
a system is always $\ket{0}$, which mathematically is [1, 0, …,
0]. This state evolves as gates (qubit transformations) are applied to
it, resulting in a final state. This process is called SV
simulation. SV simulation encounters three significant challenges at
scale:

\begin{itemize}

\item Memory-Bound Roofline Model: With Operational Intensity (OI)
  below 0.5~\cite{H_ner_2017}, SV simulation involves strided memory
  accesses during matrix-vector multiplication. Traditional cache
  hierarchies struggle with such ``strided'' memory accesses. DiaQ
  addresses this challenge by transforming strided memory accesses
  into linear accesses through different data storage techniques.
  
\item Communication Hurdles: When a SV exceeds the capacity of a
  single node, it needs to be split and stored across multiple
  nodes. Communication between these nodes can become a major
  bottleneck for simulation performance.

\item Computation Efficiency: This aspect revolves around the question
  of whether SV simulation can leverage heterogeneous accelerators to
  enhance simulation speed.
  
\end{itemize}

\textbf{Density-Matrix (DM) Simulation:}
Density matrices provide a comprehensive representation of the quantum
state, capturing both pure and mixed states. Unlike SV simulations,
density-matrix simulations accommodate quantum systems with
entanglement and statistical mixtures. The density matrix has
dimensions \((2^n, 2^n)\), and it characterizes the quantum state's
statistical information. Density-matrix simulations are particularly
advantageous when dealing with noisy quantum systems or situations
where classical uncertainty is involved. However, the computational
demands for density-matrix simulations grow quadratically with the
number of qubits, i.e., \(\mathcal{O}(2^{2*n})\). Addressing this
scalability challenge is crucial for efficiently simulating large
quantum systems and exploring noisy intermediate-scale quantum (NISQ)
algorithms. We expect our DiaQ format to significantly speed-up DM
simulation.

\textbf{Unitary Simulation:}
Every quantum gate is unitary in nature (i.e., its conjugate transpose
equals its inverse). A quantum circuit can also be represented as a
single unitary matrix with dimensions ($2^{n}$, $2^{n}$). The
generation of the quantum circuit's unitary using individual gate
unitaries is called unitary simulation. The memory requirements of
unitaries double with every addition of a qubit to the system, i.e.,
the unitary dimensions grow exponentially $\mathcal{O}(2^{n})$. And if
we were to combine two quantum circuits (unitary times unitary), a
dense algorithm for such matrix multiplication is of the order of
$\mathcal{O}(2^{3*n})$. Our DiaQ format significantly speeds up
unitary simulations, but there are few practical needs for large
unitary simulations.

\textbf{Tensor-Network (TN) Simulation:}
The size of the SV grows exponentially,  limiting simulation to the
compute node's RAM. Tensor networks are smart factorizations intended
for such large vectors, because a gate is limited to a factor of the
previously-huge vector, ''contracting'' it with another tensor (the gate
tensor has a rank equal to the number of qubits it transforms), which
reduces computation at the expense of accuracy.

This work primarily focuses on state-vector simulation. While the DiaQ
format significantly improves performance and memory efficiency of
unitary simulation, unitary simulation has limited practical
applications of relevance. Hence, our focus centers on its impact on
SV simulation, which is highly relevant for quantum computing using
classical computers as well as quantum device verification. We
contribute to the development of algorithmic advantages made possible
by the DiaQ format. In addition, density-matrix and tensor networks
simulations may provide future avenues of investigation for DiaQ
benefits, which are beyond the scope of this work.

\subsection{Benchmark Circuits} 

In this work, we consider a total of 14 quantum algorithms as
benchmarks, three from SupermarQ~\cite{supermarq} and eleven from
QASMBench~\cite{li2022qasmbench}.

The algorithms from SupermarQ are Greenberger-Horne-Zeilinger (GHZ),
Hamiltonian Simulation and Mermin-Bell. The GHZ Benchmark assesses a
quantum processor's entanglement generation using CNOT ladders to
create a GHZ state, with evaluation based on Hellinger
fidelity. Mermin-Bell tests quantumness through a Mermin-Bell
inequality, involving the preparation of a GHZ state and measuring a
specific operator's expectation value. Hamiltonian Simulation targets
the simulation of the 1D Transverse Field Ising Model (TFIM) for N
spins using Trotterization, measuring average magnetization against
classical expectations. We use this suite for scaling up the number of
qubits.

\begin{figure}[htb]
  \centering
  \subfigure[HAM(2) State]{
    \includegraphics[width=0.3\textwidth]{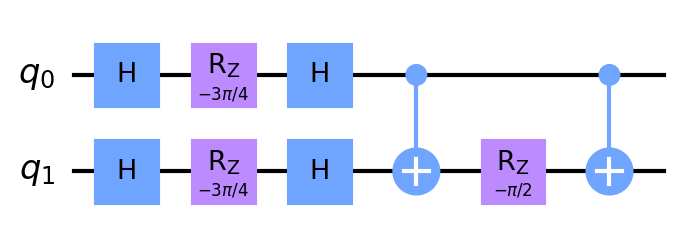}
    \label{fig:ham_state}
  }\hfill
  \subfigure[Timesteps in HAM(2) without gate parallelism]{
    \includegraphics[width=0.4\textwidth]{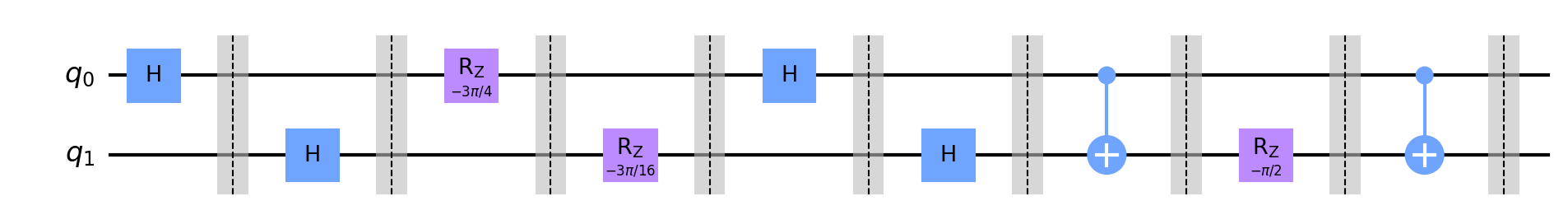}
    \label{fig:ham_state_timesteps}
  }\hfill
  \subfigure[Circuits representing various sparsity patterns (from left to right): $H \otimes I_{8}$, $I_{2} \otimes H \otimes I_{4}$, $I_{4} \otimes H \otimes I_{2}$, and $I_{8} \otimes H$.]{
    \includegraphics[width=0.1\textwidth]{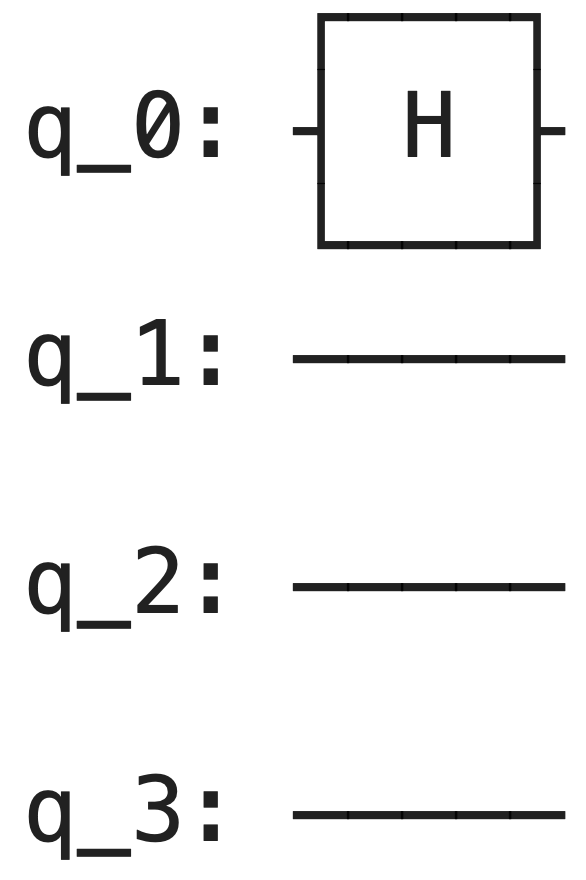}
    \includegraphics[width=0.1\textwidth]{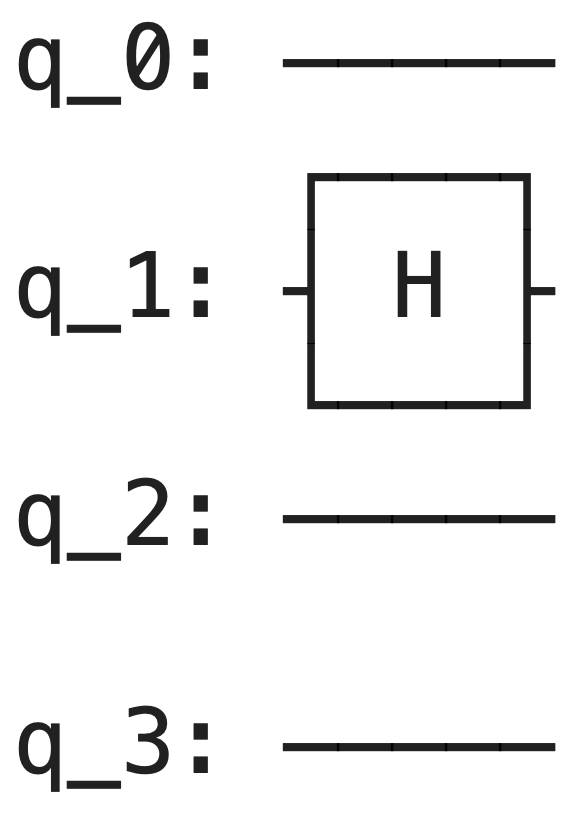}
    \includegraphics[width=0.1\textwidth]{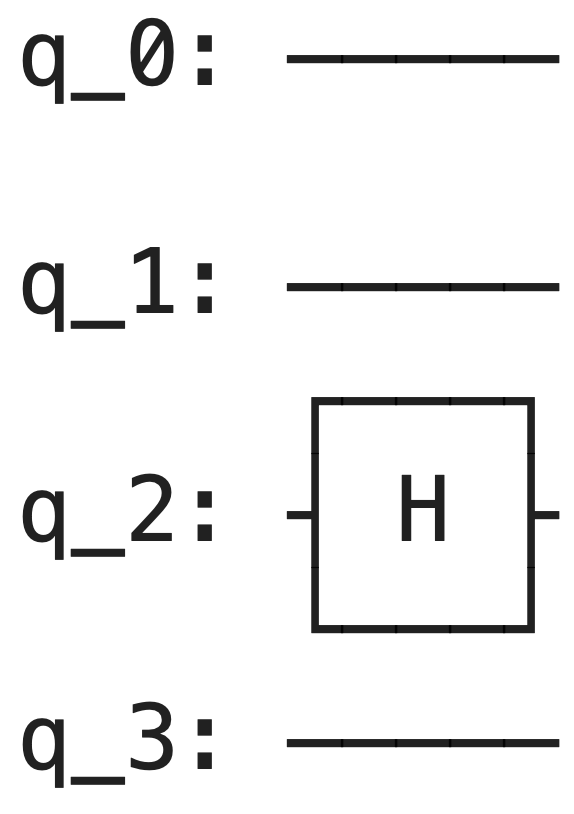}
    \includegraphics[width=0.1\textwidth]{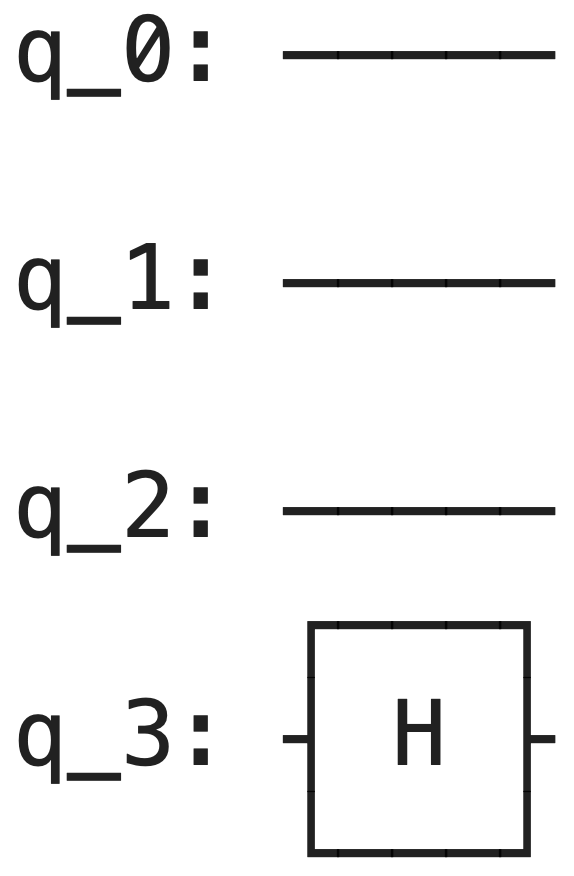}
    \label{fig:hadamard_sparsity_patterns}
  }
  \caption{Timesteps in Quantum Circuits}.
  \label{fig:timesteps_in_qc}
\end{figure}

\begin{figure*}[t]
  \subfigure[{$H \otimes I_{8}$}]
  {
    \includegraphics[width=0.22\textwidth]{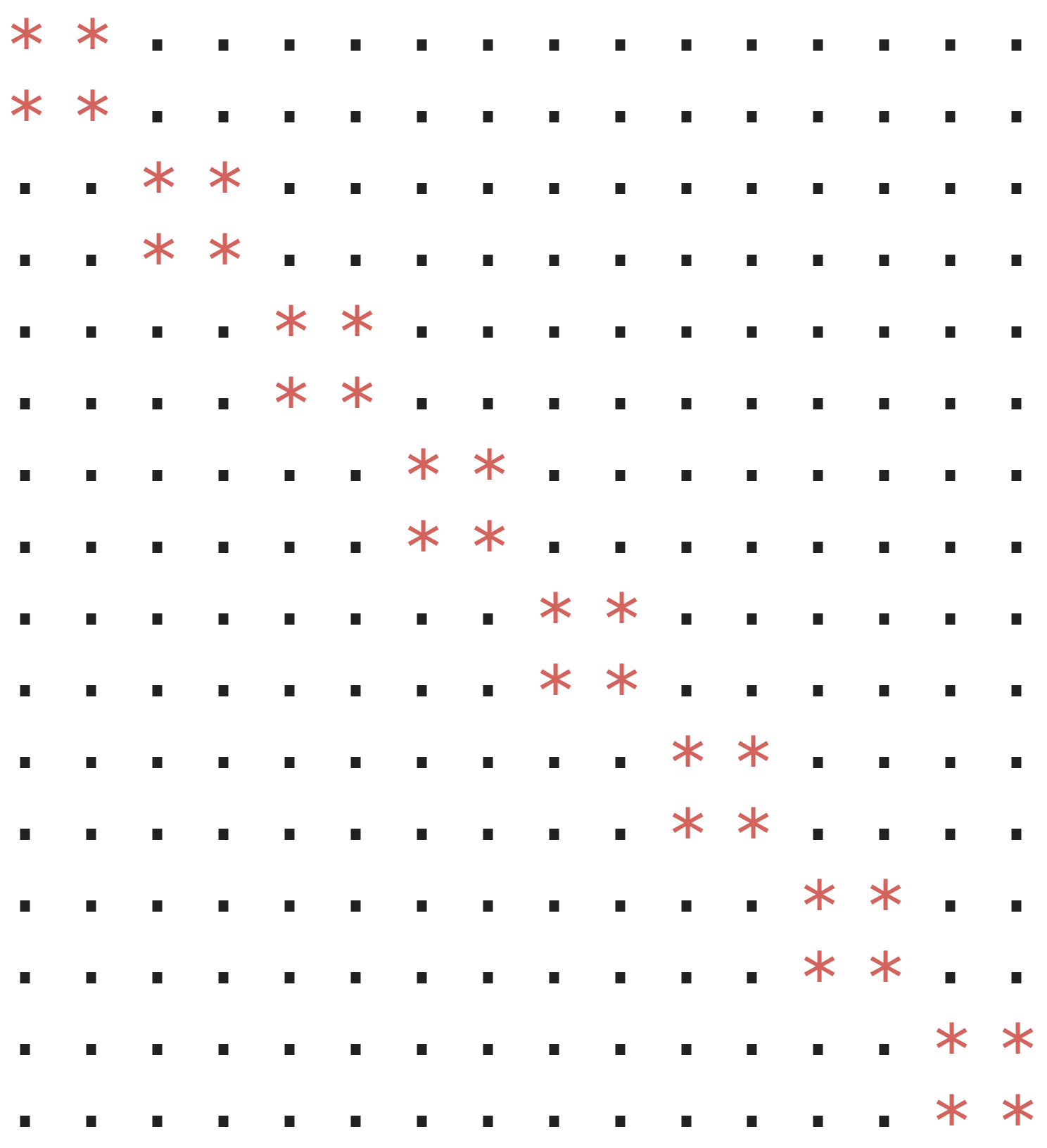}
    \label{fig:pattern_1}
  }
  \hfill
  \subfigure[{$I_{2} \otimes H \otimes I_{4}$}]
  {
    \includegraphics[width=0.22\textwidth]{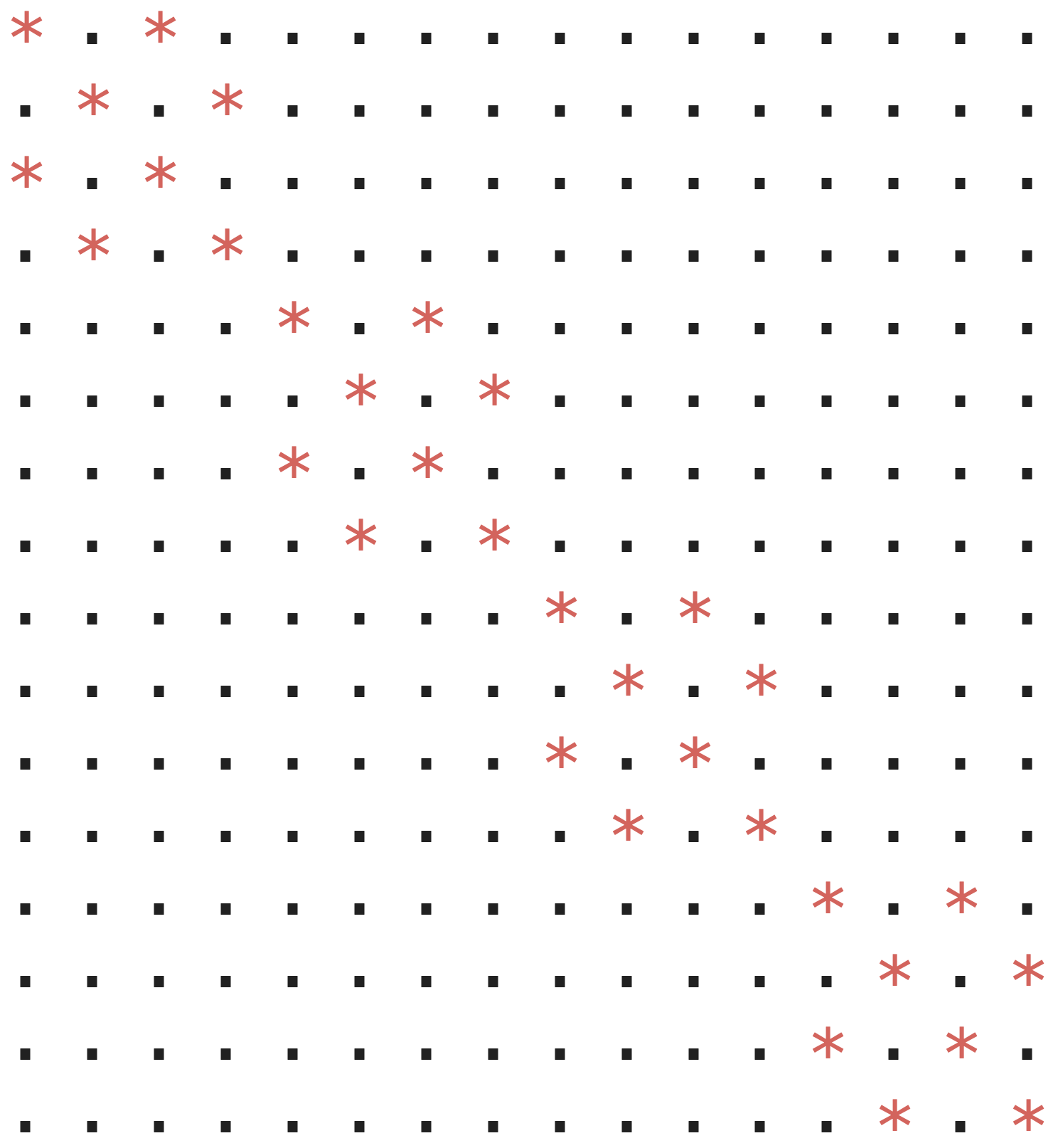}
    \label{fig:pattern_2}
  }
  \hfill
  \subfigure[{$I_{4} \otimes H \otimes I_{2}$}]
  {
    \includegraphics[width=0.22\textwidth]{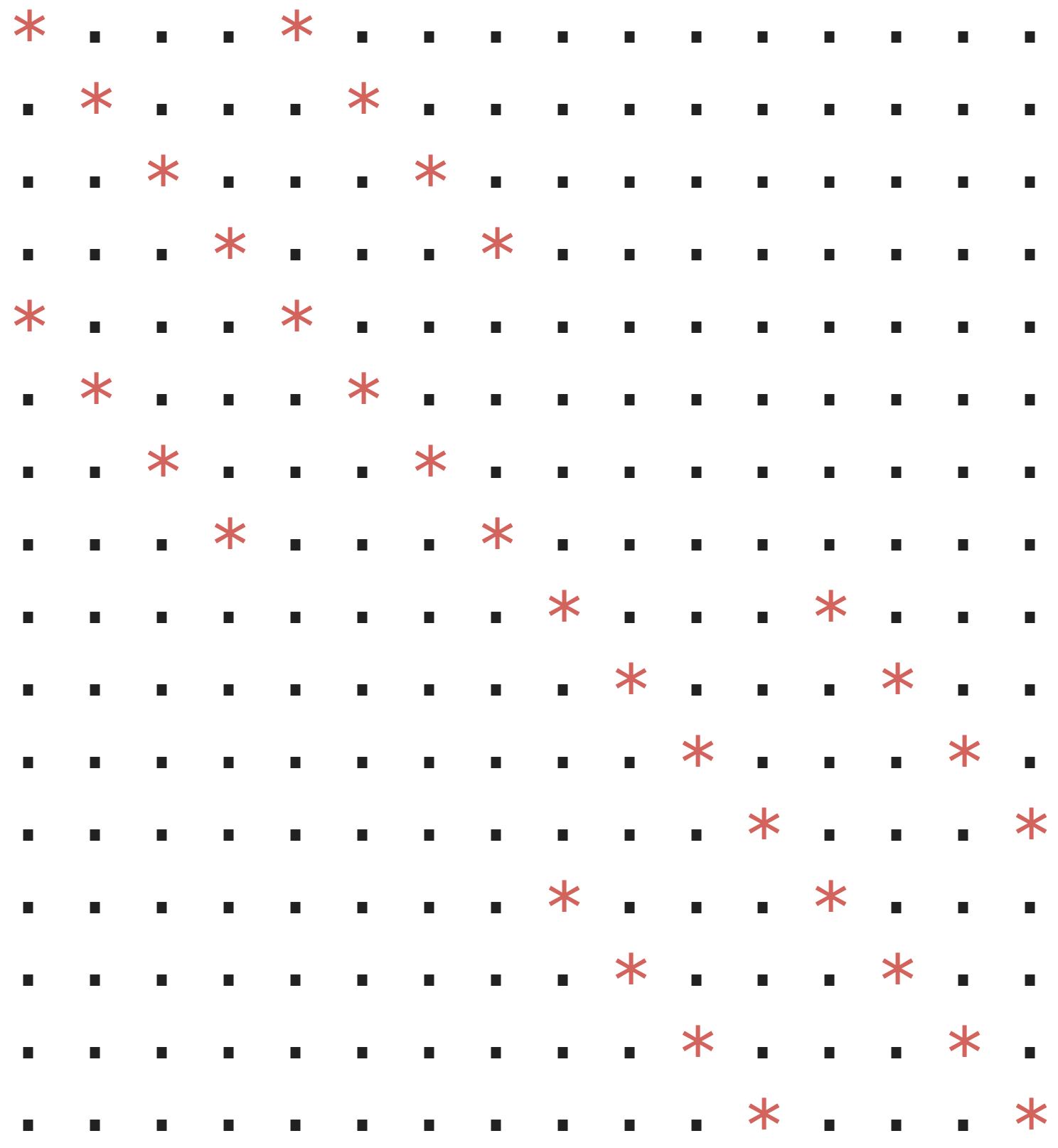}
    \label{fig:pattern_3}
  }
  \hfill
  \subfigure[{$I_{8} \otimes H$}]
  {
    \includegraphics[width=0.22\textwidth]{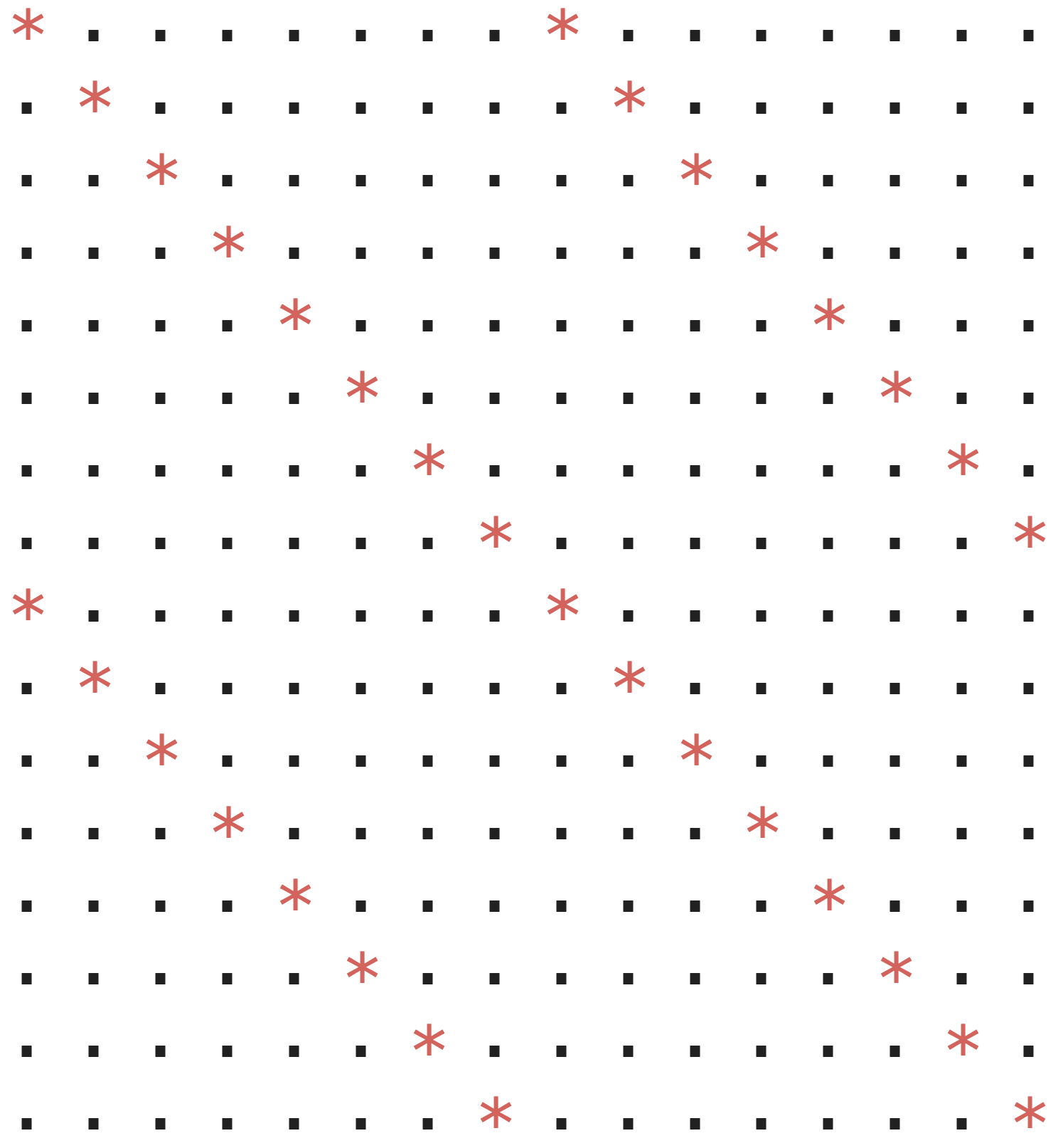}
    \label{fig:pattern_4}
  }
  \caption{Sparsity patterns of time-step unitaries in a 4-qubit
    circuit when a single Hadamard gate is applied to just the (a)
    first, (b) second, (c) third and (d), or the last qubit. Unitaries represent circuits from Figure~\ref{fig:hadamard_sparsity_patterns}. Note: The red stars denote non-zeros, black dots denote zeros.}
  \label{fig:sparsity_patterns}
  \vspace*{-\baselineskip}
\end{figure*}

The other eleven quantum algorithms that are from QASMBench~\cite{li2022qasmbench} represent
circuits covering diverse quantum algorithms and applications,
including quantum Fourier transform (\texttt{qft}), Ising model
(\texttt{ising}), secret sharing (\texttt{seca}), 3×5 multiplication
(\texttt{multiplier}), Bernstein-Vazirani algorithm (\texttt{bv}),
W-state generation (\texttt{w\_state}), larger Bernstein-Vazirani
algorithm (\texttt{bv}), counterfeit-coin finding (\texttt{cc}),
quantum ripple-carry adder (\texttt{bigadder}), quantum RAM
(\texttt{qram}), quantum Fourier transform (\texttt{qft}), adder
(\texttt{adder}), Bernstein-Vazirani algorithm (\texttt{bv}), and
quantum phase estimation to factor 21 (\texttt{qf21}). Each routine in
both sets is characterized by the number of qubits, gates, and CX
(CNOT) category. We use this benchmark suite to facilitate a direct
comparison with SV-Sim for a more comprehensive evaluation.

\subsection{Time-Steps in Quantum Circuits}

Since quantum circuits can be modeled at as a sequence of gates that
are applied one after the other, we refer to the point in time when
the gate is applied as a ``time-step''. For instance, for the quantum
circuit in Figure~\ref{fig:ham_state}, the timesteps (without gate
parallelism) are represented in Figure~\ref{fig:ham_state_timesteps},
separated by logical barriers to illustrate the sequence of steps.

The unitary at each time-step can be expressed as \( I_{\text{above}}
\otimes \text{Gate} \otimes I_{\text{below}} \), where \(
I_{\text{above}} \) and \( I_{\text{below}} \) are identity matrices
representing the state of qubits at logical positions above and below
the target qubit(s) that the gate is applied to.

\begin{figure*}[t]
  \centering {
    \includegraphics[width=0.3\textwidth]{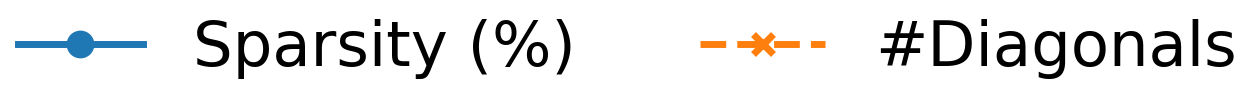}
  } \\
  \subfigure[{GHZ time-step unitaries}]
  {
    \includegraphics[width=0.3\textwidth]{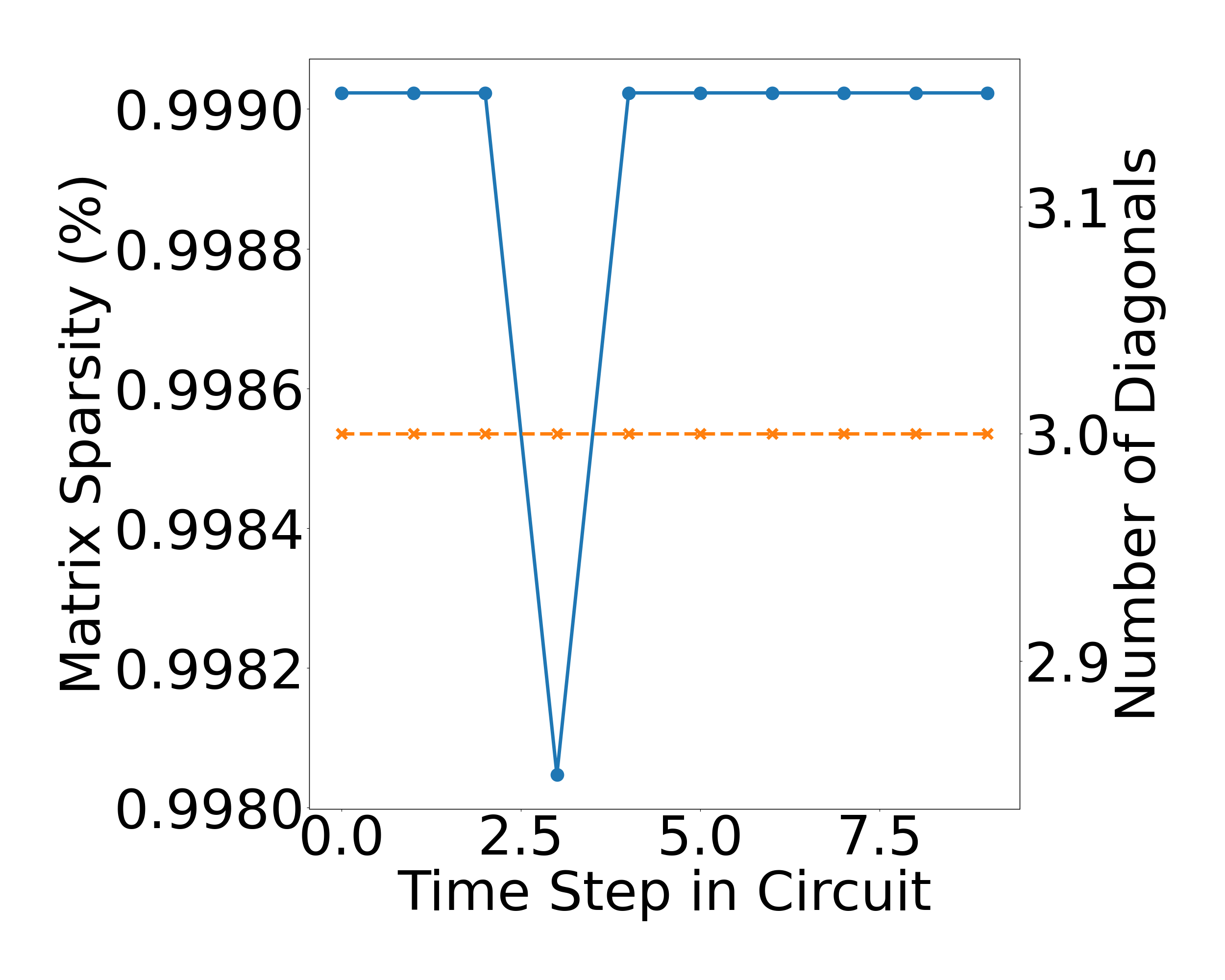}
    \label{fig:ghz10_circuit_matrices}
  }
  \hfill
  \subfigure[{Hamiltonian time-step unitaries}]
  {
    \includegraphics[width=0.3\textwidth]{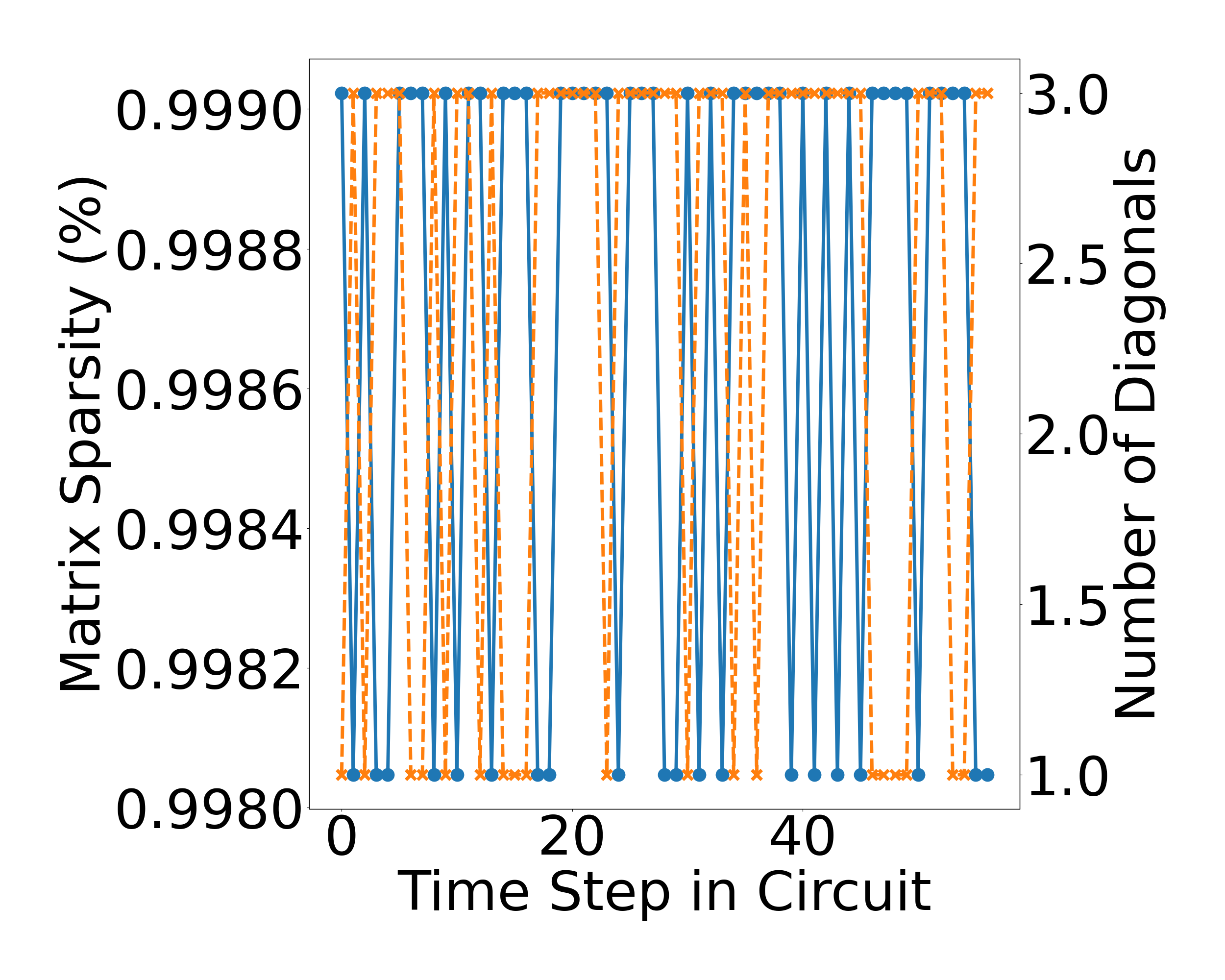}
    \label{fig:ham10_circuit_matrices}
  }
  \hfill
  \subfigure[{Mermin Bell time-step unitaries}]
  {
    \includegraphics[width=0.3\textwidth]{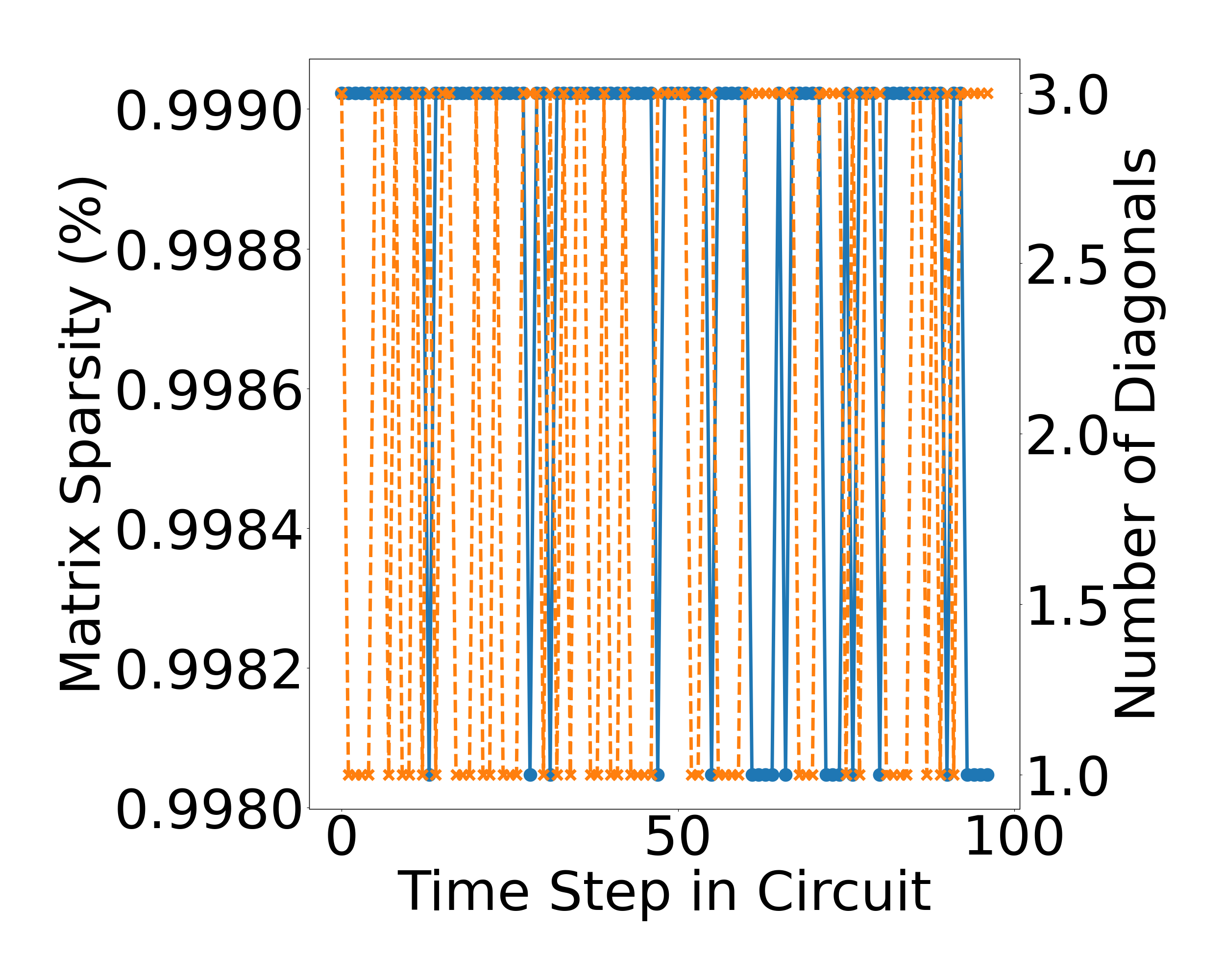}
    \label{fig:mermin_bell10_circuit_matrices}
  }
  
  \subfigure[{GHZ during unitary simulation}]
  {
    \includegraphics[width=0.3\textwidth]{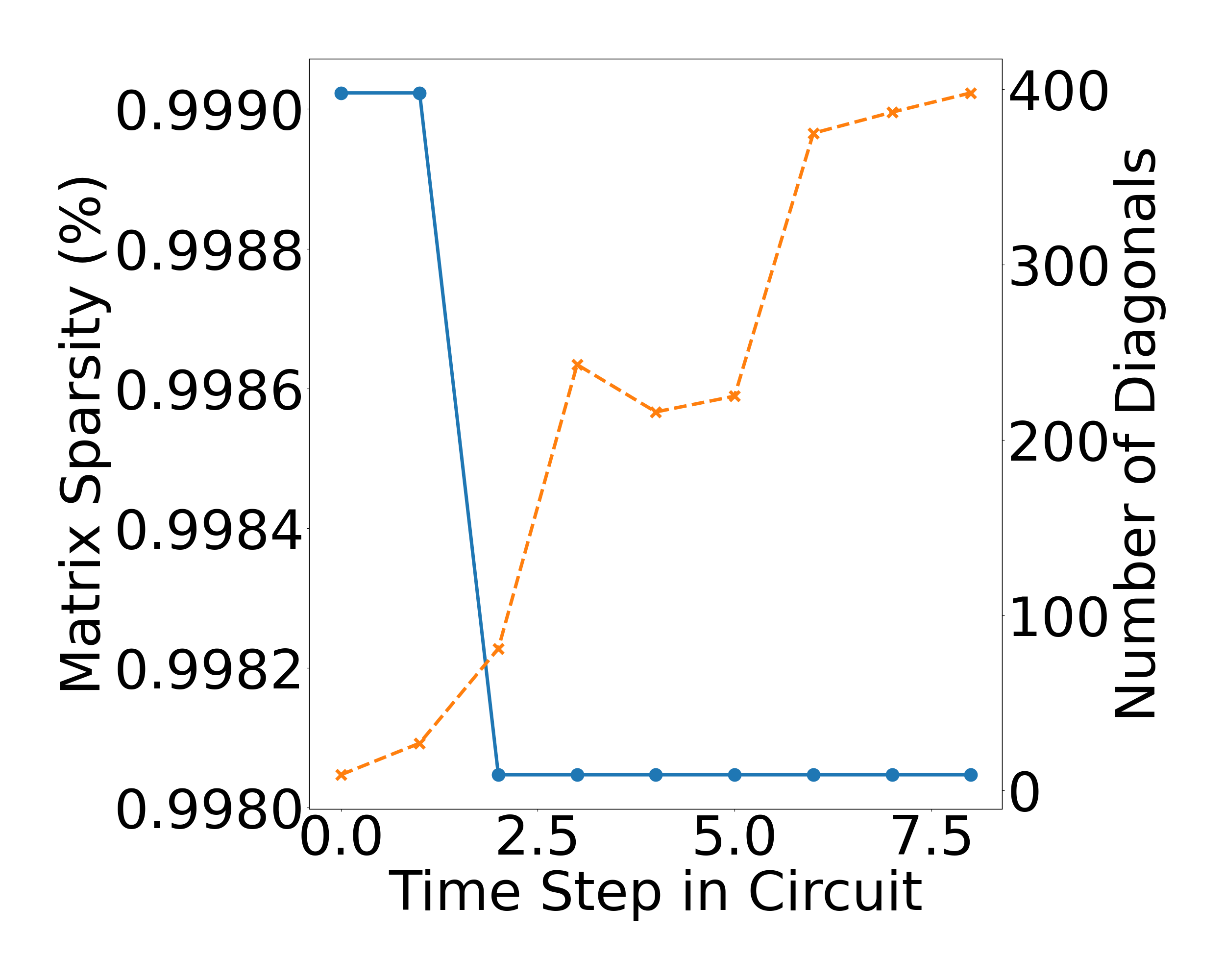}
    \label{fig:unitary_simulation_ghz10_sparsity}
  }
  \hfill
  \subfigure[{Sparsity Analysis of HAM during unitary simulation}]
  {
    \includegraphics[width=0.3\textwidth]{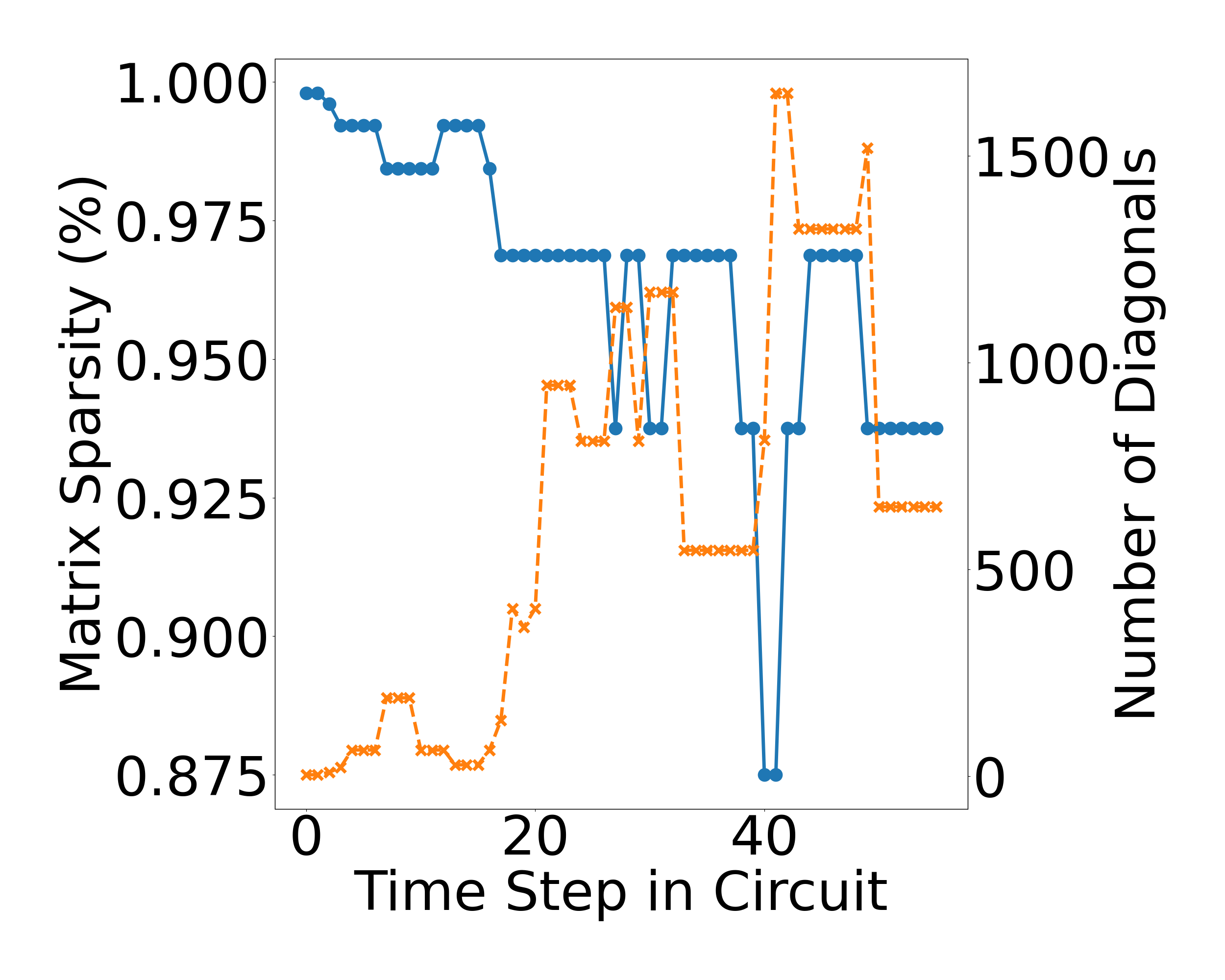}
    \label{fig:unitary_simulation_ham10_sparsity}
  }
  \hfill
  \subfigure[{Sparsity Analysis of Mermin\_Bell during unitary simulation}]
  {
    \includegraphics[width=0.3\textwidth]{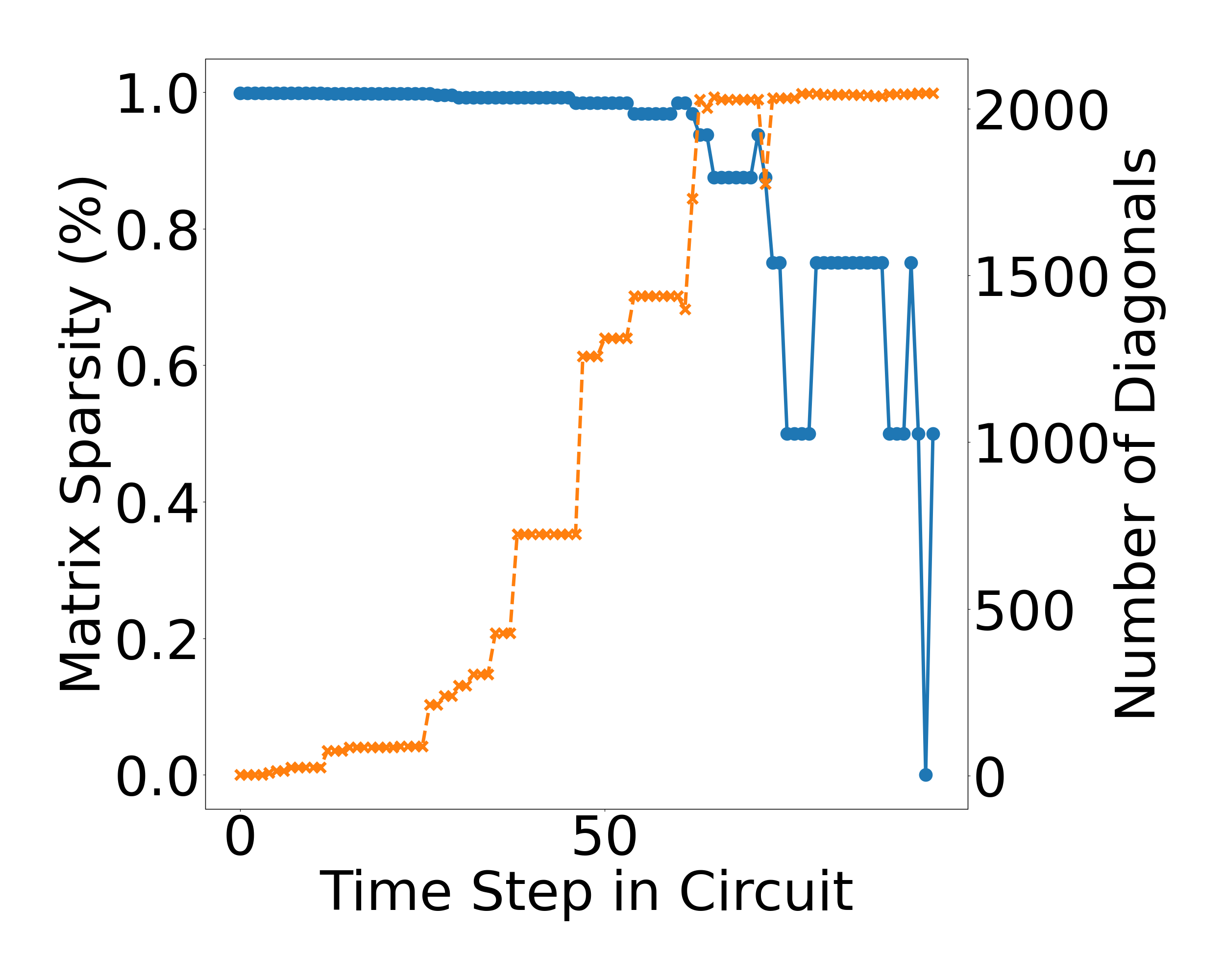}
    \label{fig:unitary_simulation_mermin_bell10_sparsity}
  }

  \caption{Sparsity in SupermarQ benchmarks}
  \label{fig:sparsity_in_supermarq_benchmarks}
  \vspace*{-\baselineskip}
\end{figure*}

\section{Motivation}

\subsection{Sparsity Patterns in QC}


Time-steps, when looked at individually as unitaries, exhibit certain
sparsity patterns. Figure~\ref{fig:sparsity_patterns} depicts
unitaries 
of time-steps where a Hadamard gate is applied to different
qubits in a 4-qubit circuit. We see a peculiar sparsity pattern: the
non-zeros in the unitaries are diagonally dense. The number of
diagonals for such unitaries are usually constant for a given quantum
gate. This is true for all norm maintaining (unitary)
transformations. In fact, for the Z (phase) gate, the unitary is just
the principal diagonal.

We find this diagonal sparsity in all time-step unitaries that are
formed from norm conserving gate
unitaries. Figures~\ref{fig:ghz10_circuit_matrices},~\ref{fig:ham10_circuit_matrices}
and~\ref{fig:mermin_bell10_circuit_matrices} indicate the number of
diagonals on the second y-axis, in time-step for unitaries of
benchmarks with 10 qubits (each unitary is a 1024x1024 matrix) from
SupermarQ~\cite{supermarq}.

Also, the primary y-axis indicates sparsity, where intermediate result
matrices during this chain matrix multiplication also are highly
sparse for certain circuits. For instance, the GHZ(10) unitary
simulation remains highly sparse ($\geq$99.8\%) throughout the
simulation as seen in
Figure~\ref{fig:unitary_simulation_ghz10_sparsity}. But some circuits
become dense after a few time-steps. For instance, the
Mermin\_Bell(10) unitary simulation becomes half-dense at time-step
72, and full-dense around time-step 98.

We hence establish that either full-circuits or sub-circuits have a
peculiar diagonal sparsity which can be exploited algorithmically for
performance gains.

There is considerable potential to improve the matrix multiplication
from a computational complexity of ($\mathcal{O}(N^3)$) to
$\mathcal{O}(N)$ when this sparsity is exploited, where $N$ is $2^{n}$.

Similarly, when there are limited diagonals in unitaries, the scope to improve the SpMV kernel, a key kernel in state-vector simulation, can also be improved, moving from $\mathcal{O}(N^2)$ to $\mathcal{O}(N * d)$, where $d$ is the number of diagonals in the matrix.

Therefore, this work leverages these sparsity patterns to formulate a
smart storage format that complements multiple linear-algebra kernels
to significantly speed up simulation and save memory.

\subsection{Existing Sparse Representations}

But before we get into our new format, let us discuss Scipy's DIA
format~\cite{2020_SciPy,saad1990sparskit,li2013smat}, which is a representation method for sparse
matrices by arranging elements in a diagonal-major pattern, employing
offsets to denote positions beyond the matrix's immediate
scope. Consider the following:

\begin{equation}\label{eq:example_matrix}
  \begin{pmatrix}
    \tikzmark{top}{a} & 0 & 0 & \tikzmark{end}{b} \\
    0 & c & 0 & 0 \\
    0 & 0 & d & 0 \\
    \tikzmark{end2}{e} & 0 & 0 & \tikzmark{bottom}{f} \\
  \end{pmatrix}
\end{equation}
\begin{tikzpicture}[overlay,remember picture]
  \draw[red,opacity=.2,line width=3mm,line cap=round] (top.center) -- (bottom.center);
  \draw[red,opacity=.2,line width=3mm,line cap=round] (end.center) -- (end.center);
  \draw[red,opacity=.2,line width=3mm,line cap=round] (end2.center) -- (end2.center);
\end{tikzpicture}

\begin{figure*}[t]
  \centering
  \includegraphics[width=0.7\textwidth]{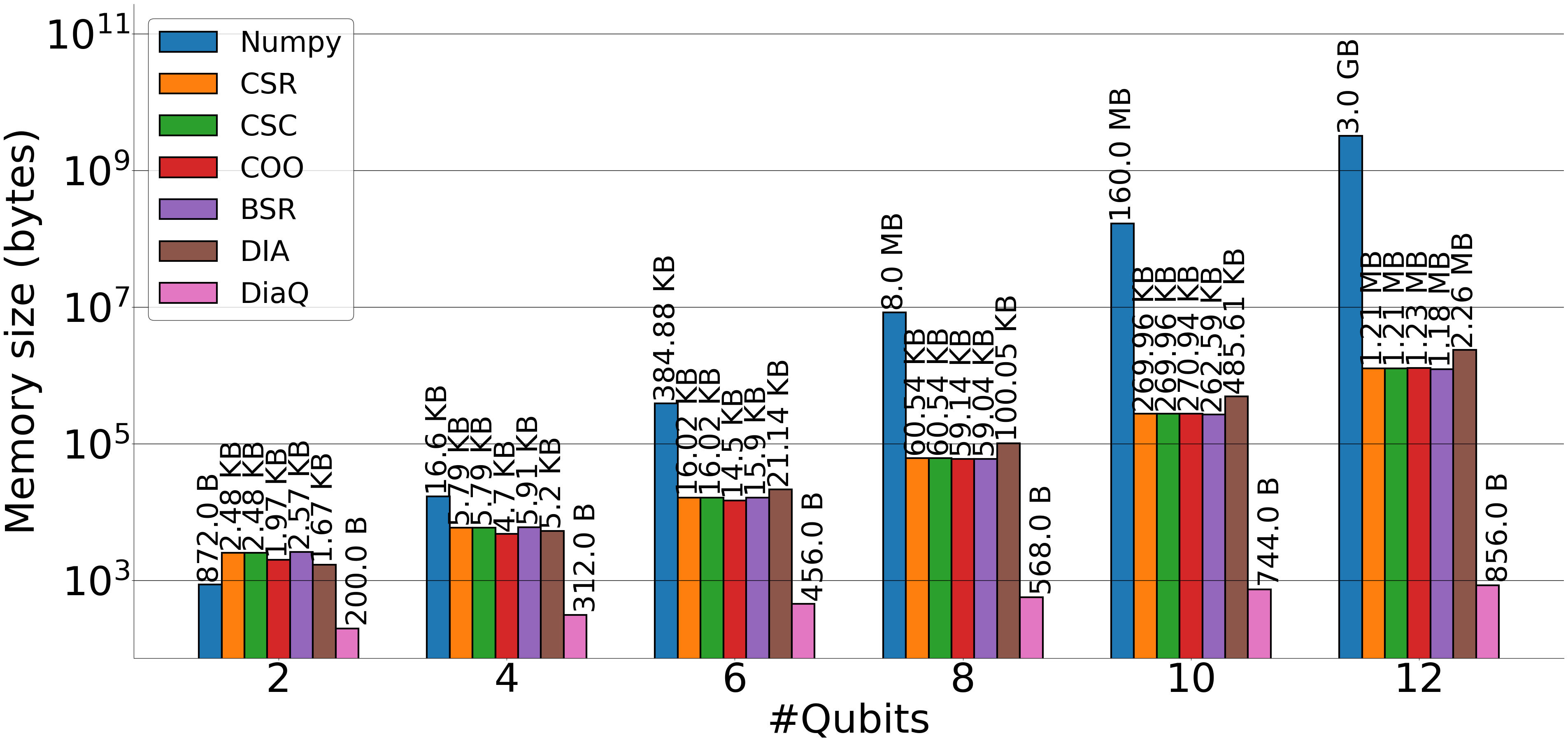}
  \vspace*{-\baselineskip}
  \caption{
    Memory Savings that DiaQ offers for GHZ circuit's chain of unitaries
  }
  \label{fig:memory_savings_diaq}
  \vspace*{-\baselineskip}
\end{figure*}

In the DIA format, non-zero elements are stored with respect to their
diagonals. For this matrix, the non-zero elements would be stored as:
\text{values} = [a, c, d, f, *, *, *, e, b, *, *, *]
Here, '*' denotes non-existent or empty positions within the matrix,
essentially representing 'NA'. The corresponding offsets would be:
\text{offsets} = [0, -3, 3]
These offsets signify the positions of the non-zero elements along the
diagonals. Specifically, the values are aligned based on their
diagonal positions relative to the main diagonal (0 offset). This
representation allows for a compact storage scheme, where non-zero
values are organized in line with their respective diagonal offsets,
effectively capturing the essential elements of the matrix in a
structured diagonal-major format.

Other commonly used sparse formats are Compressed Sparse Row (CSR),
Compressed Sparse Column (CSC), Coordinate (COO), and Block Compressed
Row (BSR) formats. CSR and CSC formats store sparse matrices by
compressing the rows or columns, respectively, utilizing arrays to
store the non-zero values and additional arrays to store the column or
row indices. On the other hand, the COO format stores the non-zero
elements along with their corresponding row and column indices, making
it suitable for constructing sparse matrices from scratch. BSR format,
unlike CSR and CSC, divides the matrix into fixed-size blocks and
compresses each block separately, enabling efficient storage and
manipulation of block-structured matrices. However, while these
formats offer efficient storage and operations for general sparse
matrices, they are not tailored specifically for quantum simulations,
where exploiting quantum-specific sparsity patterns can lead to
further performance improvements. Scipy's DIA format relies on CSR
kernels after a mid-conversion step to perform efficient matrix
operations, ensuring compatibility with existing numerical libraries
and computational environments. \\
The value array is the same for all the formats, i.e., [a, c, d, f, b,
e]. The CSR format stores Row Indices: [0, 1, 2, 3, 0, 3] and Column
Pointers: [0, 2, 3, 4, 6]. The CSC format stores Column Indices: [0,
2, 3, 3, 0, 0] and Row Pointers: [0, 1, 2, 3, 4, 6]. The COO format
stores Row Indices: [0, 1, 2, 3, 3, 0] and Column Indices: [0, 2, 3,
3, 0, 3]. The BSR format stores Block Indices: [0, 0, 1, 1, 2, 2] and
Row Indices within blocks: [0, 1, 0, 1, 0, 1].

\section{DiaQ: A Novel Quantum-Tailored Format}

Levering the sparsity patterns seen in the above section, we formulate
DiaQ, a DIA-like matrix format without the need to store offsets. We
store a hashmap of diagonals indexed by diagonal indices and values as
arrays of lengths that are pre-computed using the diagonal index and
shape of the matrix.
\\
{
  \hspace{-10pt}
  \fbox{\texttt{data[diagonal index] \(\leftarrow\) diagonal elements}}
}
\\
The above example Matrix(\ref{eq:example_matrix}) when stored in DiaQ looks like: \\
\[
  \boxed{
    \begin{aligned}
      & (dIndex=-3) & & [e] \\
      & (dIndex=0)  & & [a, c, d, f] \\
      & (dIndex=3)  & & [b]
    \end{aligned}
  }
\]

This way of storing non-zeros provides significant memory-savings for
matrices that most quantum simulations
witness. Figure~\ref{fig:memory_savings_diaq} shows DiaQ's
memory-savings compared to other sparse formats to store GHZ's chain
of unitaries for different number of qubit circuits. The figure shows
that the memory requirements of Numpy's dense format exhibit
exponential growth, whereas other sparse formats, though more compact
than Numpy, still fall short of the efficiency achieved by
DiaQ. Notably, DiaQ's memory usage scales linearly with the number of
qubits, offering substantial savings in memory
utilization. Additionally, most sparse formats necessitate conversion
to CSR before undergoing matrix operations, whereas DiaQ circumvents
this step. This exceptional scalability of DiaQ, coupled with its
efficient algorithms discussed below, establishes it as the optimal
choice for quantum simulations. Certain matrix kernels' performance
directly impacts overall simulation efficiency, necessitating a
reconsideration of BLAS operations with DiaQ matrices in mind. This
approach would optimize memory usage and computational efficiency,
potentially enhancing overall simulation performance.

\textbf{Matrix Product:}
This kernel holds significant importance as it facilitates the ``fusion
of gates" in state-vector simulations, mixed-state evolution (density
matrix simulation), and unitary simulations. The traditional approach
to Dense Matrix Product involves matrices of dimensions \(N \times
N\), with a computational complexity of \(\mathcal{O}(N^3)\). While
this dense kernel has undergone several optimizations in the realm of
high-performance computing, such as blocking, SIMD, and GPU
acceleration, it does not offer memory savings similar to
DiaQ. Storing matrices as sets of diagonals offers benefits to the
matrix product, as it can be decomposed into a series of sub-kernels
(called \texttt{MultiplyDiagonals} that perform diagonal times
diagonal computations). If both matrices contain only the principal
diagonal (diagonal index = 0), the matrix product reduces to
\(\mathcal{O}(N)\) for DiaQ as opposed to $N^2$ unitaries with
$\mathcal{O}(N^3)$ for dense matrix multiplication, where \(N\) is the
matrix size. Overall, the time complexity of this kernel is
\(\mathcal{O}(d1 * d2 * N)\), where \(d1\) and \(d2\) represent the
number of diagonals in the matrices. Each of these sub-kernels
operates in \(\mathcal{O}(N)\) time, as the diagonal length can be at
most \(N\), the dimension of the matrix. This algorithm, outlined in
Algorithm~\ref{alg:matrix_multiplication}, in the worst-case scenario,
when the matrix is fully dense, \(d1 = d2 = N\), is comparable to the
dense version, albeit complicating caching opportunities such as
blocking.

\begin{algorithm}[htb]
  \caption{Matrix Multiplication ($A \times B$)}\label{alg:matrix_multiplication}
  \begin{algorithmic}[1]
    \If{\(\text{A.columns} \neq \text{B.rows}\)}
      \State \textbf{throw} \text{InvalidArgument}(``not multiplyable")
    \EndIf
    \State \(\text{result} \gets \) \text{InitializeResultMatrix}()
    \State \(\text{mapA} \gets\) \text{A.getDiagonalMap}()
    \State \(\text{mapB} \gets\) \text{B.getDiagonalMap}()
    \State \textcolor{red}{\#pragma omp parallel for}
    \For{\(\text{diagA} \in \text{mapA}\)}
      \For{\(\text{diagB} \in \text{mapB}\)}
        \State \(\text{diagNew} \gets\) \text{MultiplyDiagonals}(\(\text{diagA}, \text{diagB}\))
        \State \Comment{\textcolor{blue}{SIMD vectorized}}
        \If{\(\text{diagNew.isValid}()\)}
          \State \(\text{newIndex} \gets \text{diagA.index + diagB.index}\)
          \State \(\text{result[newIndex]} \gets \text{diagNew}\)
        \EndIf
      \EndFor
    \EndFor
    \State \textbf{return} \(\text{result}\)
  \end{algorithmic}
\end{algorithm}

\textbf{Matrix times Vector:}
This kernel is pivotal in state-vector simulations and profoundly
impacts their performance. The use of matrices in the DiaQ format
allows for a linear time complexity in the SpMV operation when the
matrix has just one diagonal, typically $\mathcal{O}(\text{num\_diags}
\times \text{vector length})$. This represents a significant
algorithmic enhancement compared to its dense counterpart, which
typically requires $\mathcal{O}(n^2)$ time, where $n$ is the size of
the matrix. Therefore, employing the DiaQ format not only improves
memory utilization but also enhances computational efficiency. In
contrast, CSR (Compressed Sparse Row) matrix times vector takes
$\mathcal{O}(\text{nnz} + \text{num\_rows})$ time, where $\text{nnz}$
represents the number of non-zero elements in the matrix. This format
is particularly efficient for sparse matrices with a structured
pattern of non-zero elements. Similarly, the COO (Coordinate) matrix
times vector operation also achieves a time complexity of
$\mathcal{O}(\text{nnz})$. The COO format excels in scenarios where
the matrix is irregular and its non-zero elements are distributed
randomly.

\begin{algorithm}[htb]
  \caption{Sparse Matrix-Vector Product ($A \times x$)}\label{alg:sparse_matrix_vector_product}
  \begin{algorithmic}[1]
    \If{$\text{A.columns} \neq \text{x.rows}$}
      \State \textbf{throw} \text{InvalidArgument}(``not multiplyable")
    \EndIf
    \State $y \gets$ \text{InitializeResultVector}$(\text{A.matrix\_shape[0]}, 0.0)$
    \State $\text{mapA} \gets \text{A}.\text{getDiagonalMap}()$
    \State \textcolor{red}{\#pragma omp parallel for}
    \For{$\text{diagA} \in \text{mapA}$}
      \State $\text{diagA\_vals} \gets \text{diag}.\text{getValues}()$
      \If{$\text{diagA.index} < 0$} \Comment{Negative diagonals}
        \For{$i \gets 0$ \textbf{to} $\text{diagA.length}-1$} \Comment{\textcolor{blue}{SIMD}}
          \State $\text{xIndex} \gets i$
          \State $\text{yIndex} \gets i - \text{diagA.index}$
          \State $y[\text{yIndex}] \mathrel{+}= \text{diagA\_vals}[i] \cdot x[\text{xIndex}]$
        \EndFor
      \ElsIf{$\text{dIndex} > 0$} \Comment{Positive diagonals}
        \For{$i \gets 0$ \textbf{to} $\text{diagA.length}-1$} \Comment{\textcolor{blue}{SIMD}}
          \State $\text{xIndex} \gets \text{diagA.index} + i$
          \State $\text{yIndex} \gets i$
          \State $y[\text{yIndex}] \mathrel{+}= \text{diagA\_vals}[i] \cdot x[\text{xIndex}]$
        \EndFor
      \Else \Comment{Principal diagonal}
        \For{$i \gets 0$ \textbf{to} $\text{diagA.length}-1$} \Comment{\textcolor{blue}{SIMD}}
          \State $y[i] \mathrel{+}= \text{diagA\_vals}[i] \cdot x[i]$
        \EndFor
      \EndIf
    \EndFor
  \State \textbf{return} $y$
\end{algorithmic}
\end{algorithm}
    
%
Matrix Transpose is another important kernel in quantum simulations
which is not used by SV-Sim but is often utilized in BLAS
kernels~\cite{blackford2002updated}. DiaQ Matrix transpose is a linear
operation where diagonal indices are multiplied by -1, effectively
swapping the upper and lower halves of the matrix. In contrast, the
dense version takes $\mathcal{O}(N^2)$ time, where N is $2^n$ for an
$n$-qubit circuit.

            
            
            
\section{The DiaQ Library}

The DiaQ library encompasses all the algorithms discussed in the
previous section, implemented in C++ and compiled with optimization
flags (-O3) to produce a shared library and an API header file. These
artifacts facilitate seamless integration with other
libraries. Additionally, Python wrappers with converters from Numpy to
DiaQ are provided.

Furthermore, the DiaQ library offers several features to enhance its
usability and performance. Users have the flexibility to opt for byte
alignment while allocating DiaQ matrices and state-vector arrays,
catering to architectures supporting AVX SIMD vectorization
instructions and OpenMP multi-core parallelization. Additionally,
users can choose between single-precision (float) and double-precision
(double) data types, aligning the library's interface with other
High-Performance Computing (HPC) numerical libraries.

The \texttt{MultiplyDiagonals} sub-kernel, utilized in
Algorithm~\ref{alg:matrix_multiplication} at line 9, performs a linear
element-by-element operation between four arrays (real and imaginary
parts of two diagonals). This critical operation has been optimized
through vectorization using AVX-512 instructions, enhancing
computational efficiency.

\begin{algorithm}[htb]
  \caption{$y \gets (I_{\text{dim}_a} \otimes M \otimes I_{\text{dim}_b}) \times x$}\label{alg:combined_algorithm}
  \begin{algorithmic}[1]
    \State $y \gets \text{InitializeResultVector}()$
    \State $y_{\text{values}} \gets y.\text{getValues}()$
    \State $x_{\text{values}} \gets \mathbf{x}.\text{getValues}()$
    \State $\text{diagMap} \gets M.\text{getDiagonalMap}()$
    \For{each $(\text{dIndex}, \text{diag})$ in $\text{diagMap}$}
      \State $\text{dLength} \gets \text{diag.dLength}$
      \State $\text{values} \gets \text{diag.getValues}()$
      \If{$\text{dIndex} < 0$} \Comment{Handle Negative Diagonals}
        \State \textcolor{red}{\#pragma omp parallel for}
        \For{$\text{rep} \gets 0$ \textbf{to} $\text{dim}_a - 1$}
          \State $\text{skip} \gets \text{rep} \times (\text{dim}_b \times (\text{dLength} - \text{dIndex}))$
          \For{$i \gets 0$ \textbf{to} $\text{dLength}-1$}
            \State $\text{val} \gets \text{values}[i]$
            \For{$j \gets 0$ \textbf{to} $\text{dim}_b - 1$} \Comment{\textcolor{blue}{SIMD}}
              \State $\text{x\_idx} \gets j + i \times \text{dim}_b + \text{skip}$
              \State $\text{y\_idx} \gets \text{x\_idx} - (\text{dim}_b \times \text{dIndex})$
              \State $y_{\text{values}}[\text{y\_idx}] \mathrel{+}= \text{val} \times x_{\text{values}}[\text{x\_idx}]$
            \EndFor
          \EndFor
        \EndFor
      \Else \Comment{Handle Principal and Positive Diagonals}
        \State \textcolor{red}{\#pragma omp parallel for}
        \For{$\text{rep} \gets 0$ \textbf{to} $\text{dim}_a - 1$}
          \State $\text{skip} \gets \text{rep} \times (\text{dim}_b \times (\text{dLength} + \text{dIndex}))$
          \For{$i \gets 0$ \textbf{to} $\text{dLength}-1$}
            \State $\text{val} \gets \text{values}[i]$
            \For{$j \gets 0$ \textbf{to} $\text{dim}_b - 1$} \Comment{\textcolor{blue}{SIMD}}
              \State $\text{y\_idx} \gets j + i \times \text{dim}_b + \text{skip}$
              \State $\text{x\_idx} \gets \text{x\_idx} + (\text{dim}_b \times \text{dIndex})$
              \State $y_{\text{values}}[\text{y\_idx}] \mathrel{+}= \text{val} \times x_{\text{values}}[\text{x\_idx}]$
            \EndFor
          \EndFor
        \EndFor
      \EndIf
    \EndFor
    \State \textbf{return} $y$
  \end{algorithmic}
\end{algorithm}

Similarly, Algorithm~\ref{alg:sparse_matrix_vector_product} benefits
from SIMD (Single Instruction, Multiple Data) vectorization for loops
at lines 9, 15, and 21. This optimization leverages parallelism to
expedite the computation of sparse matrix-vector products,
particularly advantageous for large-scale simulations.

\begin{figure*}[t]
  \centering
  \subfigure[Broadwell]{
    \includegraphics[width=0.45\textwidth]{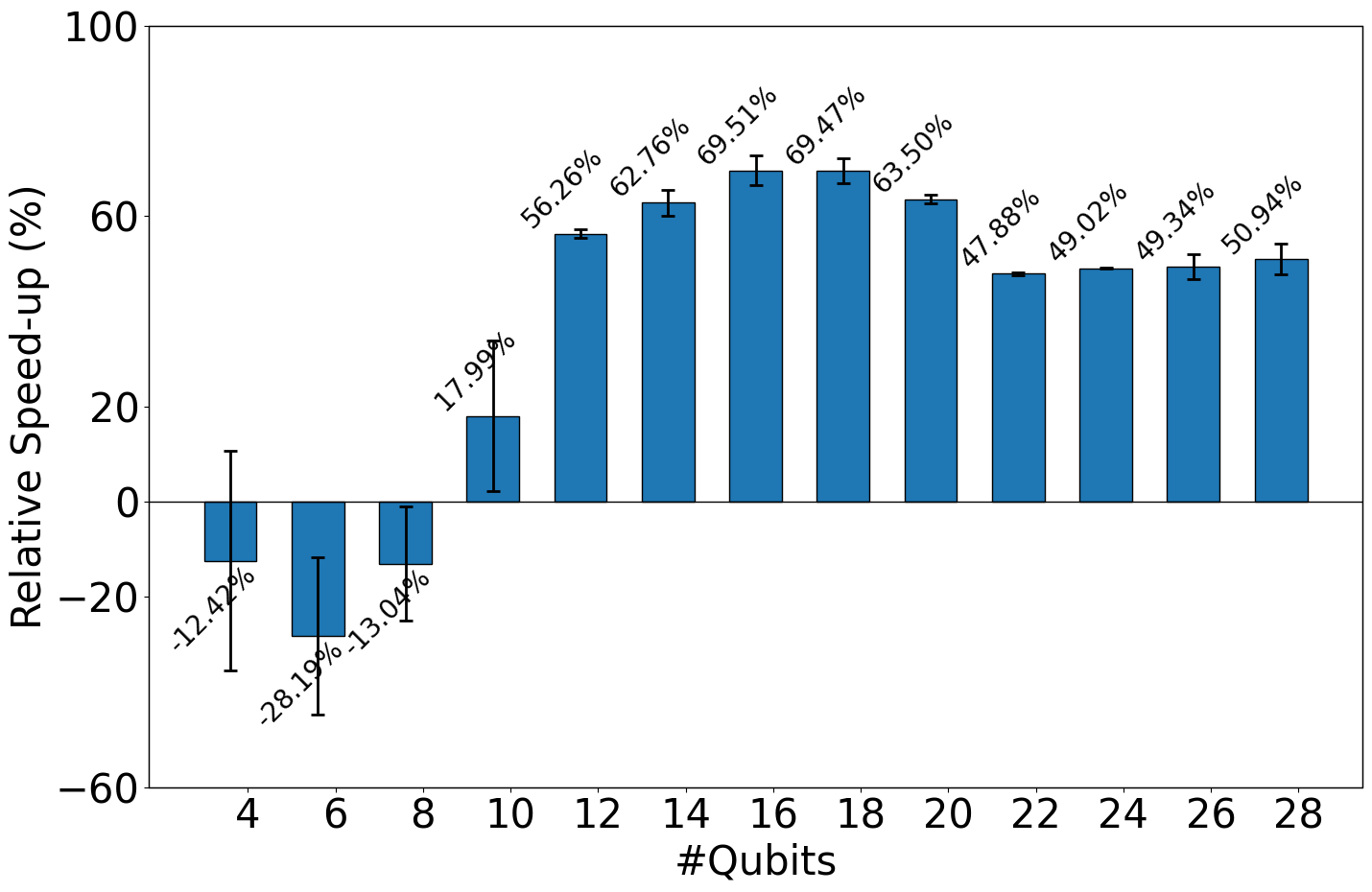}
    \label{fig:ghz_analysis}
  }
  \hfil
  \subfigure[Frontier]{
    \includegraphics[width=0.45\textwidth]{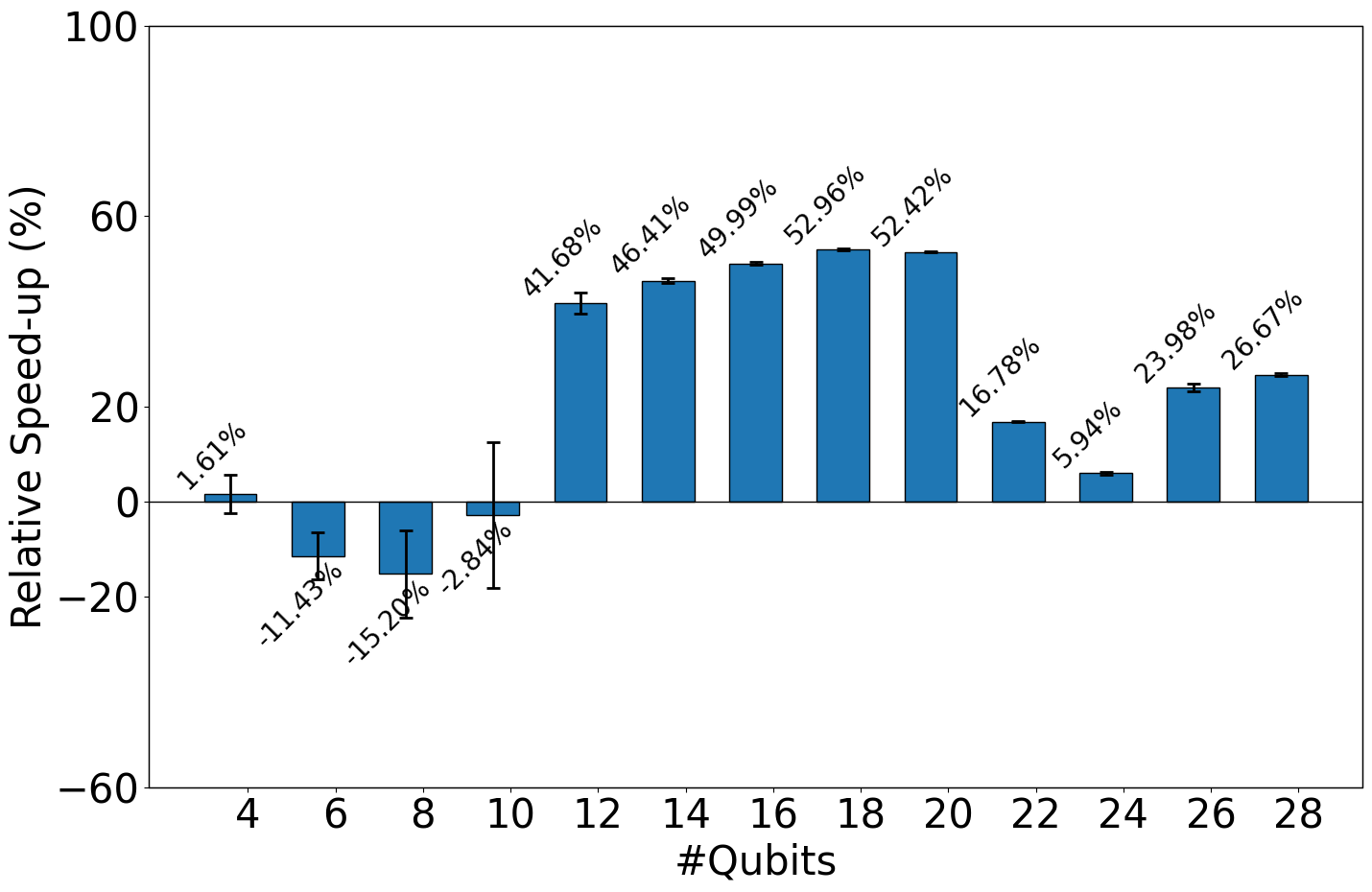}
    \label{fig:frontier_ghz_analysis}
  }
  \caption{GHZ Analysis}
  \label{fig:combined_ghz_analysis}
  \vspace*{-\baselineskip}
\end{figure*}

In Algorithm~\ref{alg:combined_algorithm}, which will be discussed in
the next section, SIMD vectorization has been applied to loops at
lines 14 and 27. This optimization further accelerates the execution
of the algorithm, contributing to overall performance enhancements.

\section{Integrating DiaQ with SV-Sim}

In this section, we discuss the integration of DiaQ, our novel sparse
matrix format, with SV-Sim, a state-vector simulator. Unlike
traditional dense representations, where gate matrices are fully
populated, DiaQ leverages sparsity by storing a hashmap of diagonal
indices to diagonal elements, resulting in efficient storage and
computation. The integration process involves utilizing sparse
Generalized Matrix-Matrix Multiplication kernel (spGEMM) for gate
fusion. Here, sparse gate matrices are efficiently multiplied to
produce a resultant sparse matrix. Moreover, gate application is
performed using sparse Matrix-Vector Multiplication kernel
(spMV). Notably, the DiaQ format allows for optimization of operations
involving matrices of the form $M \otimes I$ and $I \otimes M$,
exploiting the diagonal structure inherent in DiaQ.

\begin{figure}[htb] 
    \centering
    \scalebox{0.55}{
    \begin{tikzpicture}[cell/.style={rectangle, draw=none, minimum width=1cm, minimum height=1cm}, round/.style={circle, draw, minimum size=0.85cm}]
    
    \begin{scope}[shift={(-3,0)}] 
        \draw (0,0) grid (4,4);
        
        \node [round, fill=red!20] at (0.5, 3.5) {a};
        \node [cell] at (1.5, 3.5) {0};
        \node [round, fill=blue!20] at (2.5, 3.5) {e};
        \node [cell] at (3.5, 3.5) {0};
        
        \node [cell] at (0.5, 2.5) {0};
        \node [round, fill=green!20] at (1.5, 2.5) {b};
        \node [cell] at (2.5, 2.5) {0};
        \node [round, fill=orange!20] at (3.5, 2.5) {f};
        
        \node [round, fill=purple!20] at (0.5, 1.5) {g};
        \node [cell] at (1.5, 1.5) {0};
        \node [round, fill=cyan!20] at (2.5, 1.5) {c};
        \node [cell] at (3.5, 1.5) {0};
        
        \node [cell] at (0.5, 0.5) {0};
        \node [round, fill=magenta!20] at (1.5, 0.5) {h};
        \node [cell] at (2.5, 0.5) {0};
        \node [round, fill=yellow!20] at (3.5, 0.5) {d};
        
        \node [align=center, font=\Large] at (2,-1) {\textbf{Gate Matrix ($G$)}};
    \end{scope}
    
    \draw [->, line width=1pt] (1.2, 1.8) -- node[midway, above] {DiaQ} (2.5, 1.8);
    
    \begin{scope}[shift={(2.7, 4.75)}]
        \draw (0,0) grid (1,2);
        \node [round, fill=purple!20] at (0.5, 1.5) {g};
        \node [round, fill=magenta!20] at (0.5, 0.5) {h};
        \node [font=\Large] at (2.8, 1) {diagonal index -2};
    \end{scope}

    \begin{scope}[shift={(2.7, 0)}]
        \draw (0,0) grid (1,4);
        \node [round, fill=red!20] at (0.5, 3.5) {a};
        \node [round, fill=green!20] at (0.5, 2.5) {b};
        \node [round, fill=cyan!20] at (0.5, 1.5) {c};
        \node [round, fill=yellow!20] at (0.5, 0.5) {d};
        \node [font=\Large] at (2.8, 2) {diagonal index 0};
    \end{scope}

    \begin{scope}[shift={(2.7, -2.75)}]
        \draw (0,0) grid (1,2);
        \node [round, fill=blue!20] at (0.5, 1.5) {e};
        \node [round, fill=orange!20] at (0.5, 0.5) {f};
        \node [font=\Large] at (2.8, 1) {diagonal index 2};
    \end{scope}

    \begin{scope}[shift={(7.5,-4)}]
        \draw (0,0) grid (1,15);
        \node [font=\large] at (0.5, 14.5) {$x_0$};
        \node [font=\large] at (0.5, 13.5) {$\vdots$};
        \node [font=\large] at (0.5, 12.5) {$x_k$};
        \node [font=\large] at (0.5, 11.5) {$\vdots$};
        \node [font=\large] at (0.5, 10.5) {$x_{k+3}$};
        \node [font=\large] at (0.5, 9.5) {$\vdots$};
        \node [font=\large] at (0.5, 8.5) {$x_{l}$};
        \node [font=\large] at (0.5, 7.5) {$\vdots$};
        \node [font=\large] at (0.5, 6.5) {$x_{l+3}$};
        \node [font=\large] at (0.5, 5.5) {$\vdots$};
        \node [font=\large] at (0.5, 4.5) {$x_{m}$};
        \node [font=\large] at (0.5, 3.5) {$\vdots$};
        \node [font=\large] at (0.5, 2.5) {$x_{m+3}$};
        \node [font=\large] at (0.5, 1.5) {$\vdots$};
        \node [font=\large] at (0.5, 0.5) {$x_{N-1}$};
        \node [align=center, font=\LARGE] at (-3,-0.5) {\textbf{$y = (I_{\text{dim}_a} \otimes M \otimes I_{\text{dim}_b}) \times x$}};
    \end{scope}

    \begin{scope}[shift={(11,-4)}]
        \draw (0,0) grid (1,15);
        \node [font=\large] at (0.5, 14.5) {$y_0$};
        \node [font=\large] at (0.5, 13.5) {$\vdots$};
        \node [font=\large] at (0.5, 12.5) {$\vdots$};
        \node [font=\large] at (0.5, 11.5) {$y_k$};
        \node [font=\large] at (0.5, 10.5) {$\vdots$};
        \node [font=\large] at (0.5, 9.5) {$\vdots$};
        \node [font=\large] at (0.5, 8.5) {$y_l$};
        \node [font=\large] at (0.5, 7.5) {$\vdots$};
        \node [font=\large] at (0.5, 6.5) {$\vdots$};
        \node [font=\large] at (0.5, 5.5) {$y_m$};
        \node [font=\large] at (0.5, 4.5) {$\vdots$};
        \node [font=\large] at (0.5, 3.5) {$\vdots$};
        \node [font=\large] at (0.5, 2.5) {$\vdots$};
        \node [font=\large] at (0.5, 1.5) {$\vdots$};
        \node [font=\large] at (0.5, 0.5) {$y_{N-1}$};
    \end{scope}

    \begin{scope}[shift={(3.7, 0)}]
        \draw [dotted, line width=2pt, red] (0, 4) -- (3.8, 9);
        \draw [dotted, line width=2pt, red] (0, 0) -- (3.8, 6);

        \draw [dotted, line width=2pt, blue] (0, 4) -- (3.8, 5);
        \draw [dotted, line width=2pt, blue] (0, 0) -- (3.8, 2);
        
        \draw [dotted, line width=2pt, green] (0, 4) -- (3.8, 1);
        \draw [dotted, line width=2pt, green] (0, 0) -- (3.8, -2);
    \end{scope}

    \begin{scope}[shift={(10.5, 2)}]
        \draw [->, line width=1pt] (-1.3, 5.5) -- (0, 5.5);
        \node at (-0.75, 5.0) {$\vdots$};
        \draw [->, line width=1pt] (-1.3, 3.5) -- (0, 3.5);
        \node at (-0.75, 4.0) {$\vdots$};
        \draw [->, line width=1pt] (-1.3, 1.5) -- (0, 1.5);
        \node at (-0.75, 3.0) {$\vdots$};
        \draw [->, line width=1pt] (-1.3, -0.5) -- (0, -0.5);
        \node at (-0.75, 2.0) {$\vdots$};
        \draw [->, line width=1pt] (-1.3, -2.5) -- (0, -2.5);
        \node at (-0.75, 1.0) {$\vdots$};
        \node at (-0.75, 0.0) {$\vdots$};
        \node at (-0.75, -1.0) {$\vdots$};
        \node at (-0.75, -2.0) {$\vdots$};
    \end{scope}

    \end{tikzpicture}
    }
    \caption{Gate Application (Algorithm \ref{alg:combined_algorithm}). In this example, we focus on just the principal diagonal (\( \text{dIndex} = 0 \)). Each element along this diagonal, denoted by \( i \), interacts with the corresponding element in the input vector \( x \). Specifically, the \( i \)th element in the diagonal is multiplied with the \( (j + i \times \text{dim}_b + \text{skip}) \)th element in the vector \( x \), where \( \text{skip} = 4 \times \text{rep} \times \text{dimB} \), iterates over the diagonal \( \text{dimA} \) times, where \( \text{rep} \) signifies each iteration and \( \text{dimB} \) represents the columns in \( I_{\text{dim}_b} \), i.e., qubits above those that the gate is applied to. And \( j \) iterates over \( \text{dim}_b \). In brief, each set of four elements in the input vector \( x \) corresponds directly to four elements in the gate matrix diagonal. The limited number of diagonals typically present in gate matrices, render this operation nearly linear.}
    \label{fig:gate_application}
\end{figure}
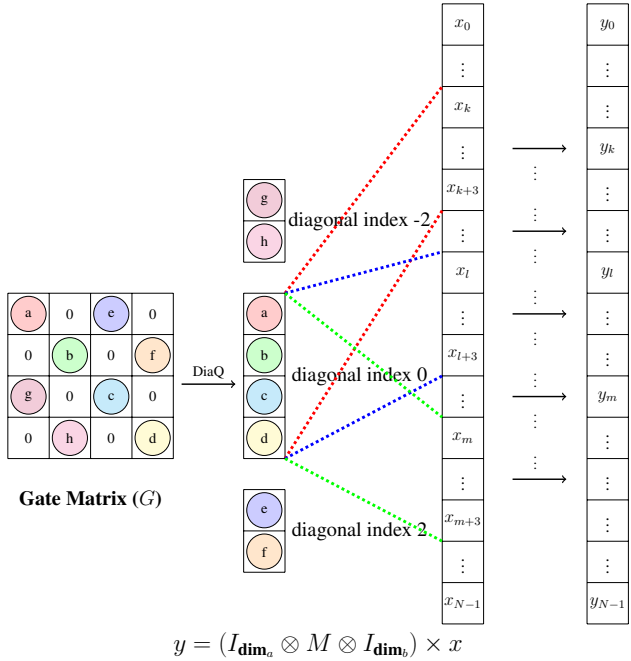

The DiaQ format's storage of data in diagonals enables contiguous
memory accesses during Sparse Matrix-Vector Multiplication
(spMV). This stands in contrast to dense representations, where memory
accesses are staggered, leading to reduced computational efficiency. A
significant enhancement arises from the nearly zero Floating Point
Operations (FLOPs) required for the Kronecker product operation,
facilitated by moving contiguous memory accesses enabled by the DiaQ
format. The reported speedups in simulations largely stem from
optimizing Kronecker product algorithms based on this
premise. Furthermore, by recognizing that Sparse Matrix-Vector
Multiplication (SpMV) operations often follow the identity Kronecker
matrix operation, we exploit this synergy to combine both operations
efficiently. This integration eliminates redundant iterations over the
same non-zero elements resulting in improved computational
performance. The details of this optimization are given in
Algorithm~\ref{alg:combined_algorithm}
and depicted in Figure~\ref{fig:gate_application}. In practice, only
diagonals (and often just the singular principle one) need to be
multiplied with shifting subsets of the state vector resulting in an
order of magnitude reduction in complexity.

\begin{figure*}[t]
  \centering
  \subfigure[Broadwell]{
    \includegraphics[width=0.45\textwidth]{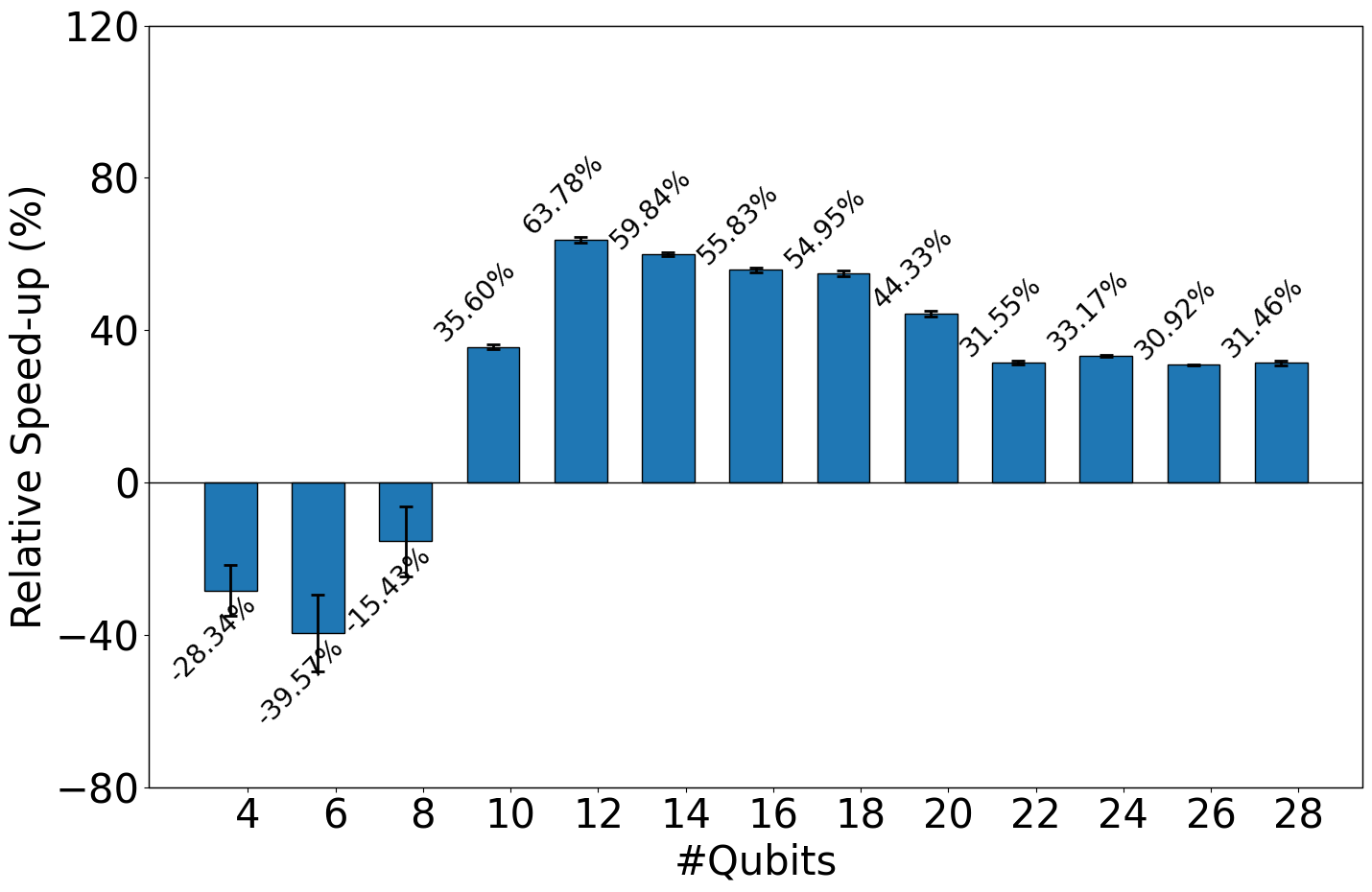}
    \label{fig:ham_analysis}
  }
  \hfil
  \subfigure[Frontier]{
    \includegraphics[width=0.45\textwidth]{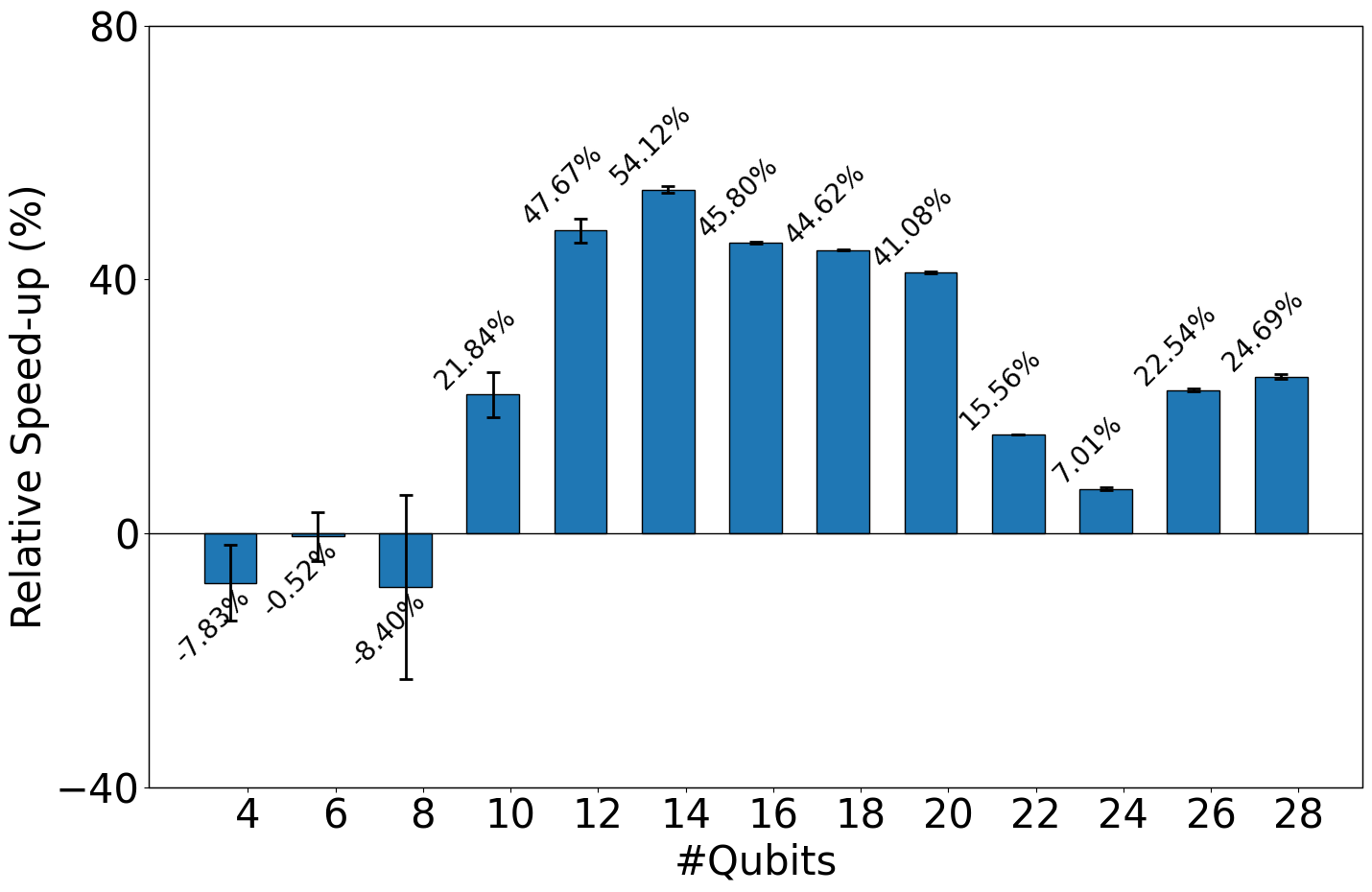}
    \label{fig:frontier_ham_analysis}
  }
  \caption{HAM Analysis}
  \label{fig:combined_ham_analysis}
  \vspace*{-\baselineskip}
\end{figure*}

The CMake configuration of the SV-Sim C++ library has been adapted to
include libdiaq.so and diaq.h from the DiaQ C++ library. This
modification allows for seamless integration with any numerical
library that provides the necessary kernels utilized by
SV-Sim. SV-Sim's backendManager is responsible for directing
simulations to different backends. diaq-cpu and diaq-mpi have been
added to the list of available backends, but at compile time.

\begin{figure*}[t]
  \centering
  \subfigure[{Broadwell}]
  {
    \includegraphics[width=0.4\textwidth]{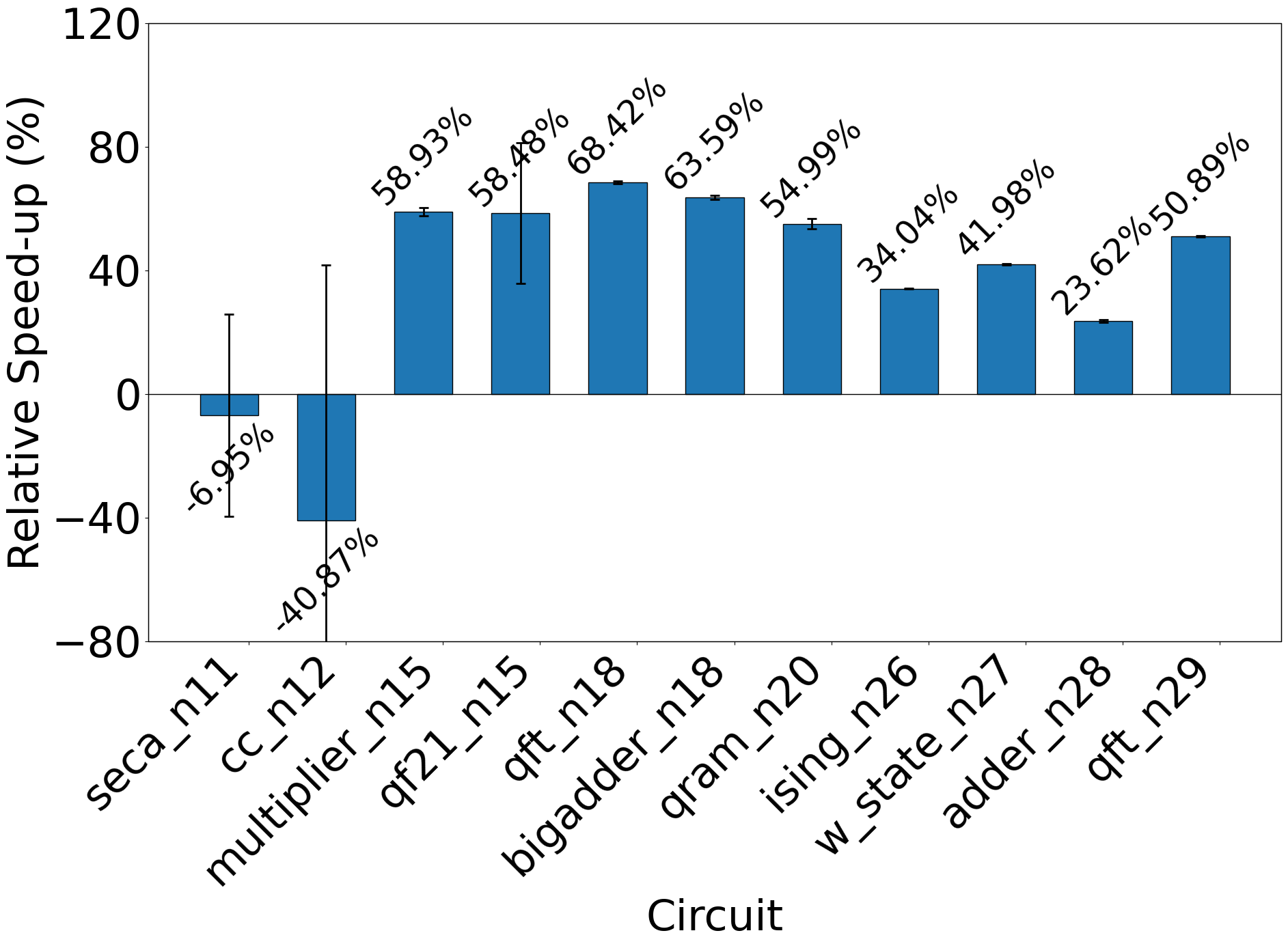}
    \label{fig:qasmbench_analysis}
  }
  \hfil
  \subfigure[{Frontier}]
  {
    \includegraphics[width=0.4\textwidth]{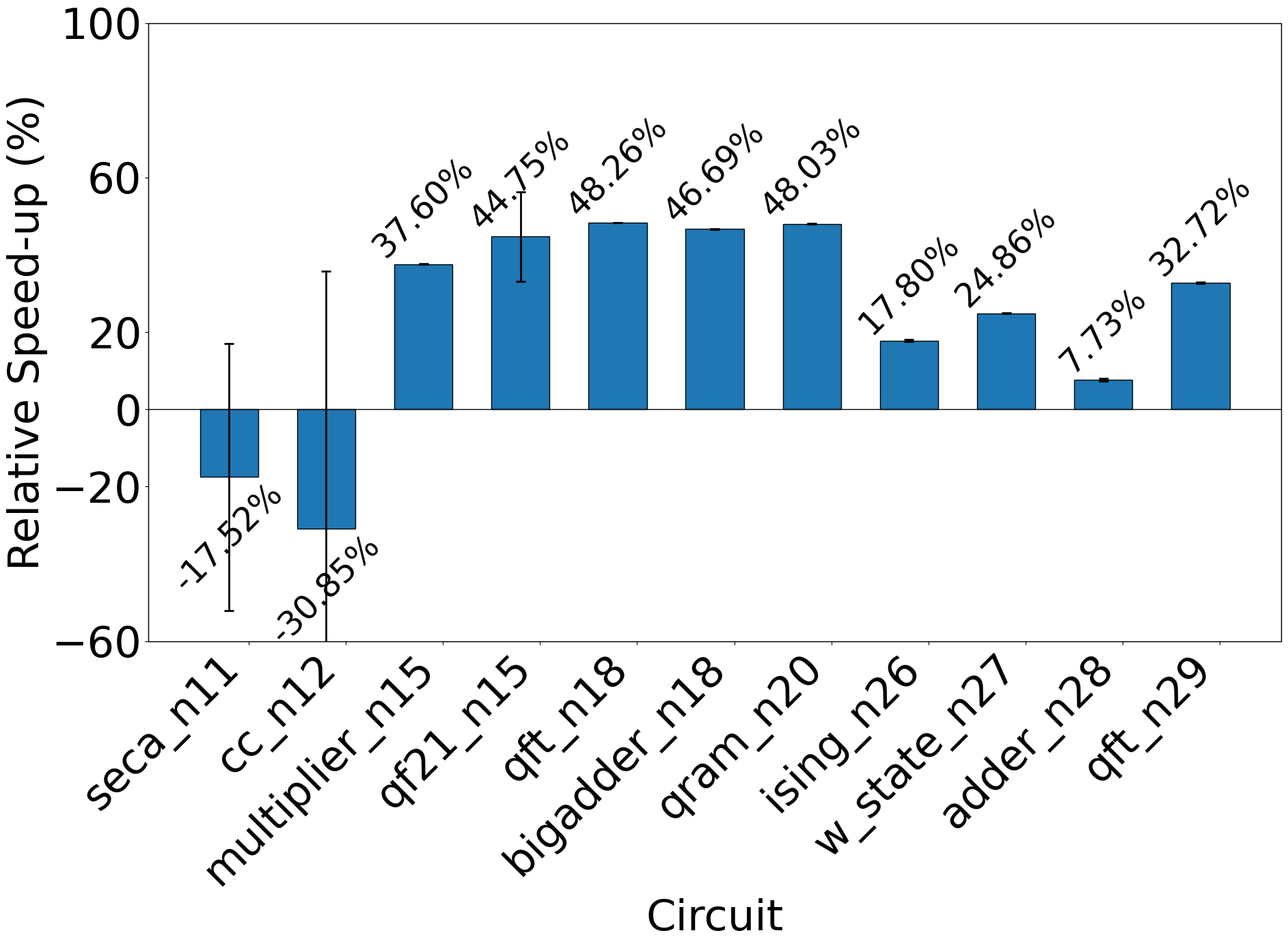}
    \label{fig:frontier_qasmbench_analysis}
  }
  
  \caption{QASMBench tests}
  \label{fig:qasmbench_comparison}
  \vspace*{-\baselineskip}
\end{figure*}

\section{Experimental Framework}
    

We convey our experiments on an Intel Broadwell Processor with 32
cores and 64GB RAM, and a Frontier~\cite{10.1145/3581784.3607089}
Cluster Node featuring 128 cores (AMD EPYC 7763) with 256GB RAM.
%

The evaluation encompassed a range of benchmarks, including GHZ and
HAM benchmarks from 2-28 qubits, from the SupermarQ suite~\cite{supermarq},
and a range of diverse quantum circuits from
QasmBench~\cite{li2022qasmbench}, such as
\texttt{seca\_n11}, \texttt{cc\_n12},
\texttt{multiplier\_n15}, \texttt{qf21\_n15}, \texttt{qft\_n18},
\texttt{bigadder\_n18},
\texttt{qram\_n20},
\texttt{ising\_n26}, \texttt{w\_state\_n27}, \texttt{adder\_n28}, and
\texttt{qft\_n29}. All benchmarks are run for 1024 shots. The averages are
plotted over 10 runs, and the box-plots on each bar show the standard
deviation over the 10 runs. Time measurements were obtained using
output of SV-Sim's trace functionality within the program.

\section{Results}

\subsection{Quantitative Results}

Figures~\ref{fig:ghz_analysis} and~\ref{fig:frontier_ghz_analysis}
depicts the relative speedup (y-axis) of our sparse DiaQ over SV-Sim's
dense baseline in the range of 4-28 qubits (x-axis) for the GHZ
benchmark on Broadwell and Frontier, respectively. We observe speedups
of the simulation runtime on average by 56-69\% and 41-52\% in the
range of 12-20 qubits on Broadwell and Frontier, respectively, where
problems fit into L3 cache. Beyond L3 capacity, speedups are around
50\% and 25\% for Broadwell (22 or more qubits) and Frontier (26 or
more qubits), respectively.
There is a dip in performance for circuits in the range of 22-24
qubits on Frontier, which can be attributed to L3 cache effects as the
overall data size just exceeds the L3 capacity, likely before hardware
prefetching becomes effective (at 26 qubits) to recover to a stable
speedup.
For small problem sizes with 8-10 or fewer qubits, the dense baseline
outperforms our sparse DiaQ due to highly optimized dense linear
algebra, which fits the hardware platforms perfectly.

Figures~\ref{fig:ham_analysis} and~\ref{fig:frontier_ham_analysis}
show the results for the HAM benchmark on the same axes, which follow
a similar trend to GHZ. We observe that the DiaQ format speeds up
runtime in its sparse DiaQ format over dense simulation for larger number
of qubits. Speedups peak around 12-14 qubits at 63\% and 54\% for
Broadwell and Frontier, respectively, and stabilize at 26 or more
qubits around 31\% and 23\% for the two respective platforms. The same
dips around 22-24 qubits are seen on Frontier, and problems sizes with
at most 8 qubits result in slowdowns for the same reasons as before.

In the following, we restrict ourselves to results of problem sizes
exceeding 10 qubits as only then can speedups be expected for DiaQ. We
omit results for Mermin Bell from the SupermarQ benchmark as this
benchmark can only construct circuits of less than 12 qubits.


Figures~\ref{fig:qasmbench_analysis}
and~\ref{fig:frontier_qasmbench_analysis} depict speedups on the
y-axis of DiaQ over the dense SV-Sim baseline for a variety of
QASMBench circuits with different number of qubits (indicated as
\_n\#\#) on the x-axis. The results show that DiaQ improves the
simulation significantly, especially for large qubit circuits with at
least 15 qubits, but for problems smaller than that slowdowns are
seen. For instance, qft at 18 qubits experiences a reduction in
simulation time by 68\% and 48\% for Broadwell and Frontier,
respectively, down to 50\% and 32\% at 29 qubits, again due to
exceeding L3 cache size. Savings for classical problems reimplemented
as a quantum circuit, such as the adder circuit, as well as state
preparation (w\_state) and Ising Hamiltonian simulation (ising), tend
to be smaller.

Overall, simulation time is significantly reduced via DiaQ, peaking at
L2 cache capacity and then stabilizing beyond L3 capacity as circuits
become larger and sparser.

\begin{table*}[!htbp]
\centering
\caption{Other Quantum Libraries}
\label{tab:quantum_libraries}
\begin{tabular}{lcccccc}
\toprule
\textbf{Library} & \textbf{MV} & \textbf{GEMM} & \textbf{CPU Multi-threaded?} & \textbf{SIMD vectorized?} & \textbf{Open-sourced?} \\
\midrule
Qiskit & $\mathcal{O}(n^2)$ & $\mathcal{O}(n^3)$ & \checkmark & \checkmark & \checkmark \\
CUDA-Quantum \cite{cuda-quantum} & $\mathcal{O}(n^2)$ & $\mathcal{O}(n^3)$ & \checkmark & -- & \texttimes \\
Cirq \cite{cirq_developers_2023_10247207} & $\mathcal{O}(n^2)$ & $\mathcal{O}(n^3)$ & \checkmark & \checkmark & \checkmark \\
Qulacs & $\mathcal{O}(n^2)$ & $\mathcal{O}(n^3)$ & -- & \checkmark & \checkmark \\
Stim & $\mathcal{O}(n^2)$ & $\mathcal{O}(n^3)$ & -- & \checkmark & \checkmark \\
SV-Sim + DiaQ (this work) & $\mathcal{O}(n \times d)$ & $\mathcal{O}(n \times d \times d)$ & \checkmark & \checkmark & \checkmark \\
\bottomrule
\end{tabular}
\begin{flushleft}
Note: In this table, $n$ represents the matrix dimension, which is $2^\text{{number of qubits}}$, and $d$ represents the number of diagonals in the matrix.
\end{flushleft}
\end{table*}

\subsection{Qualitative Analysis}

DiaQ significantly reduces memory requirements in unitary simulations,
converting $\mathcal{O}(N^2)$ to $\mathcal{O}(d \times N)$, where d if
often a constant in quantum simulations (see
Figure~\ref{fig:gate_application}). Storing in this format is
beneficial for unitary simulation as it reduces $\mathcal{O}(N^3)$
matrix multiplication to a $\mathcal{O}(d \times d \times N)$ kernel
as seen in Algorithm~\ref{alg:matrix_multiplication}, i.e., often
$\mathcal{O}(N)$ if $d$ is constant.

Most significantly, DiaQ improves state-vector simulation when the
circuit unitaries are sufficiently sparse. It achieves this by
leveraging the \texttt{spMV} kernel demonstrated in
Algorithm~\ref{alg:sparse_matrix_vector_product}, which exhibits a
complexity of $\mathcal{O}(N \times d)$, whereas the dense
Matrix-Vector product takes $\mathcal{O}(N^2)$.

As demonstrated in Algorithm~\ref{alg:combined_algorithm}, DiaQ allows
for smart optimizations to avoid intermediate Kronecker product
calculations. The complexity of the kernel
\texttt{$(I_{\text{dim}_a} \otimes M \otimes I_{\text{dim}_b}) \times
  x$} is $\mathcal{O}(N \times d)$, whereas the same sequence of steps
(Kronecker followed by Matrix-Vector) in the dense format would take
$\mathcal{O}(N^4)$.



\section{Related Work}

In the field of quantum computing, various high-performance computing
(HPC) tools have been developed for quantum simulation.
However, developing simulations for distributed and parallel systems
using classical simulators on the HPC systems requires advanced skill
in programming and algorithm design.  Benchmark tests of
QASMBench~\cite{li2022qasmbench} were conducted on the OLCF Summit
supercomputer equipped with IBM Power9 CPUs, Nvidia Volta V100 GPUs,
and a Mellanox EDR 100 Gb/s Infiniband interconnect, utilizing the
distributed NWQ-Sim simulator~\cite{suhandli2023}.  This research
leveraged significant computational resources and advanced simulation
technologies.

QSim~\cite{quantum_ai_team_and_collaborators_2020_4023103} by Google
is an open-source simulator designed for Schr{\"o}dinger
simulations. Quimb~\cite{Gray2018} utilizes tensor networks for
high-performance circuit simulation. Qulacs~\cite{Suzuki_2021} by
Quansys is another open-source simulator tailored for Schr{\"o}dinger
simulations. Stim~\cite{Gidney_2021}, an open-source library by
Google, focuses on high-speed simulation of Clifford circuits and
quantum error correction. cuQuantum~\cite{bayraktar2023cuquantum}
provides tools by Nvidia for accelerating quantum simulation on GPUs.

Recently, Nvidia introduced Quantuloop~\cite{quantuloop}, a sparse
simulator that utilizes a bitwise representation for efficient
simulation of quantum circuits on GPU architectures. Despite these
advancements, sparse simulation techniques have not been extensively
explored in the field.

DiaQ distinguishes itself from existing approaches as depicted in
Table~\ref{tab:quantum_libraries}. In particular, DiaQ specifically
targets {\em sparsity in quantum simulations at the numerical format
level}, both in unitary matrices and during vector space simulation via
algorithmic kernel specialization. This unique characteristic enables
DiaQ to offer superior scalability and efficiency compared to
traditional simulators.

The sparse simulator~\cite{jaques2021leveraging} under the Azure
Quantum Development Kit focuses on leveraging state-vector sparsity to
save memory and improve simulation time, which is orthogonal to DiaQ's
approach of exploiting sparsity in quantum simulations at the unitary
matrix level.

Moreover, DiaQ's versatility extends beyond quantum simulation to
other applications encountering diagonal sparsity, such as after
Kaleidoscope~\cite{kaleidoscope} matrices are formed or when weight
matrices are predominantly diagonal in deep neural networks. This
adaptability positions DiaQ as a promising solution for optimizing
memory usage and computational performance across various domains.

\section{Conclusion}

DiaQ represents a significant advancement in sparse matrix formats,
particularly in the realm of quantum simulation. By exploiting
diagonal sparsity, DiaQ offers substantial reductions in memory usage
and computational complexity, making it a valuable tool for quantum
researchers and practitioners. SV-Sim with DiaQ speeds up most
benchmark circuits significantly ($\sim23.65\%$) across 11 diverse
benchmarks from QASMBench via multi-core parallelization (OpenMP) and
vectorization (SIMD). DiaQ can be used like other numerical libraries,
after a simple python import.

Exploring DiaQ's compatibility and efficacy with other matrix-based
simulation techniques, such as unitary simulations and density-matrix
simulations, holds promise for further advancements and should be a
focus of future investigations. Moreover, once libdiaq gains GPU
acceleration support, it can further enhance quantum simulations. DiaQ
can be tested in conjunction with tensor-network simulations once
diagonals are re-imagined for higher-order tensors (rank$>$2). By
continuing to innovate in the realm of sparse simulation techniques,
we can unlock new opportunities for advancing quantum computing and
computational science as a whole.

\section*{\textbf{Acknowledgment}}

This work was supported in part by NSF CISE-2217020, CISE-2316201,
PHY-1818914, PHY-2325080.
This research used resources of the Oak Ridge Leadership
Computing Facility at the Oak Ridge National Laboratory, which is
supported by the Office of Science of the U.S. Department of Energy
under Contract No. DE-AC05-00OR22725.
{\it Notice}: This manuscript has been authored by UT-Battelle, LLC, under contract DE-AC05-00OR22725 with the US Department of Energy (DOE). The US government retains and the publisher, by accepting the article for publication, acknowledges that the US government retains a nonexclusive, paid-up, irrevocable, worldwide license to publish or reproduce the published form of this manuscript, or allow others to do so, for US government purposes. DOE will provide public access to these results of federally sponsored research in accordance with the DOE Public Access Plan (https://www.energy.gov/doe-public-access-plan).

\bibliography{references}

\begin{thebibliography}{10}
\providecommand{\url}[1]{#1}
\csname url@samestyle\endcsname
\providecommand{\newblock}{\relax}
\providecommand{\bibinfo}[2]{#2}
\providecommand{\BIBentrySTDinterwordspacing}{\spaceskip=0pt\relax}
\providecommand{\BIBentryALTinterwordstretchfactor}{4}
\providecommand{\BIBentryALTinterwordspacing}{\spaceskip=\fontdimen2\font plus
\BIBentryALTinterwordstretchfactor\fontdimen3\font minus
  \fontdimen4\font\relax}
\providecommand{\BIBforeignlanguage}[2]{{%
\expandafter\ifx\csname l@#1\endcsname\relax
\typeout{** WARNING: IEEEtran.bst: No hyphenation pattern has been}%
\typeout{** loaded for the language `#1'. Using the pattern for}%
\typeout{** the default language instead.}%
\else
\language=\csname l@#1\endcsname
\fi
#2}}
\providecommand{\BIBdecl}{\relax}
\BIBdecl

\bibitem{Shor_1997}
\BIBentryALTinterwordspacing
P.~W. Shor, ``Polynomial-time algorithms for prime factorization and discrete
  logarithms on a quantum computer,'' \emph{{SIAM} Journal on Computing},
  vol.~26, no.~5, pp. 1484--1509, oct 1997. [Online]. Available:
  \url{https://doi.org/10.1137%2Fs0097539795293172}
\BIBentrySTDinterwordspacing

\bibitem{anschuetz2018variational}
E.~R. Anschuetz, J.~P. Olson, A.~Aspuru-Guzik, and Y.~Cao, ``Variational
  quantum factoring,'' 2018.

\bibitem{farhi2014quantum}
E.~Farhi, J.~Goldstone, and S.~Gutmann, ``A quantum approximate optimization
  algorithm,'' 2014.

\bibitem{Peruzzo_2014}
\BIBentryALTinterwordspacing
A.~Peruzzo, J.~McClean, P.~Shadbolt, M.-H. Yung, X.-Q. Zhou, P.~J. Love,
  A.~Aspuru-Guzik, and J.~L. O'Brien, ``A variational eigenvalue solver on a
  photonic quantum processor,'' \emph{Nature Communications}, vol.~5, no.~1,
  jul 2014. [Online]. Available: \url{https://doi.org/10.1038%2Fncomms5213}
\BIBentrySTDinterwordspacing

\bibitem{Low_2019}
\BIBentryALTinterwordspacing
G.~H. Low and I.~L. Chuang, ``Hamiltonian simulation by qubitization,''
  \emph{Quantum}, vol.~3, p. 163, jul 2019. [Online]. Available:
  \url{https://doi.org/10.22331%2Fq-2019-07-12-163}
\BIBentrySTDinterwordspacing

\bibitem{Uvarov_2020}
\BIBentryALTinterwordspacing
A.~V. Uvarov, A.~S. Kardashin, and J.~D. Biamonte, ``Machine learning phase
  transitions with a quantum processor,'' \emph{Physical Review A}, vol. 102,
  no.~1, jul 2020. [Online]. Available:
  \url{https://doi.org/10.1103%2Fphysreva.102.012415}
\BIBentrySTDinterwordspacing

\bibitem{Biamonte_2017}
\BIBentryALTinterwordspacing
J.~Biamonte, P.~Wittek, N.~Pancotti, P.~Rebentrost, N.~Wiebe, and S.~Lloyd,
  ``Quantum machine learning,'' \emph{Nature}, vol. 549, no. 7671, pp.
  195--202, sep 2017. [Online]. Available:
  \url{https://doi.org/10.1038%2Fnature23474}
\BIBentrySTDinterwordspacing

\bibitem{Harrow_2009}
\BIBentryALTinterwordspacing
A.~W. Harrow, A.~Hassidim, and S.~Lloyd, ``Quantum algorithm for linear systems
  of equations,'' \emph{Physical Review Letters}, vol. 103, no.~15, oct 2009.
  [Online]. Available: \url{https://doi.org/10.1103%2Fphysrevlett.103.150502}
\BIBentrySTDinterwordspacing

\bibitem{Woerner_2019}
\BIBentryALTinterwordspacing
S.~Woerner and D.~J. Egger, ``Quantum risk analysis,'' \emph{npj Quantum
  Information}, vol.~5, no.~1, feb 2019. [Online]. Available:
  \url{https://doi.org/10.1038%2Fs41534-019-0130-6}
\BIBentrySTDinterwordspacing

\bibitem{Braine_2021}
\BIBentryALTinterwordspacing
L.~Braine, D.~J. Egger, J.~Glick, and S.~Woerner, ``Quantum algorithms for
  mixed binary optimization applied to transaction settlement,'' \emph{{IEEE}
  Transactions on Quantum Engineering}, vol.~2, pp. 1--8, 2021. [Online].
  Available: \url{https://doi.org/10.1109%2Ftqe.2021.3063635}
\BIBentrySTDinterwordspacing

\bibitem{analog_qc_material_science}
J.~Tangpanitanon, S.~Thanasilp, M.-A. Lemonde, and D.~Angelakis, ``Quantum
  supremacy with analog quantum processors for material science and machine
  learning,'' 06 2019.

\bibitem{Feynman1982Simulating}
\BIBentryALTinterwordspacing
R.~Feynman, ``Simulating physics with computers,'' \emph{International Journal
  of Theoretical Physics}, vol.~21, no. 6-7, pp. 467--488, Jun. 1982. [Online].
  Available: \url{http://dx.doi.org/10.1007/bf02650179}
\BIBentrySTDinterwordspacing

\bibitem{li2021svsim}
A.~Li, B.~Fang, C.~Granade, G.~Prawiroatmodjo, B.~Hein, M.~Rotteler, and
  S.~Krishnamoorthy, ``Sv-sim: Scalable pgas-based state vector simulation of
  quantum circuits,'' in \emph{Proceedings of the International Conference for
  High Performance Computing, Networking, Storage and Analysis}, 2021.

\bibitem{H_ner_2017}
\BIBentryALTinterwordspacing
T.~Häner and D.~S. Steiger, ``0.5 petabyte simulation of a 45-qubit quantum
  circuit,'' in \emph{Proceedings of the International Conference for High
  Performance Computing, Networking, Storage and Analysis}, ser. SC
  ’17.\hskip 1em plus 0.5em minus 0.4em\relax ACM, Nov. 2017. [Online].
  Available: \url{http://dx.doi.org/10.1145/3126908.3126947}
\BIBentrySTDinterwordspacing

\bibitem{supermarq}
\BIBentryALTinterwordspacing
T.~Tomesh, P.~Gokhale, V.~Omole, G.~S. Ravi, K.~N. Smith, J.~Viszlai, X.-C. Wu,
  N.~Hardavellas, M.~R. Martonosi, and F.~T. Chong, ``Supermarq: A scalable
  quantum benchmark suite,'' 2022. [Online]. Available:
  \url{https://arxiv.org/abs/2202.11045}
\BIBentrySTDinterwordspacing

\bibitem{li2022qasmbench}
A.~Li, S.~Stein, S.~Krishnamoorthy, and J.~Ang, ``Qasmbench: A low-level
  quantum benchmark suite for nisq evaluation and simulation,'' \emph{ACM
  Transactions on Quantum Computing}, 2022.

\bibitem{2020_SciPy}
\BIBentryALTinterwordspacing
{Virtanen, Pauli and Gommers, Ralf and Oliphant, Travis E. and Haberland, Matt
  and Reddy, Tyler and Cournapeau, David and Burovski, Evgeni and Peterson,
  Pearu and Weckesser, Warren and Bright, Jonathan and van der Walt, Stéfan J.
  and Brett, Matthew and Wilson, Joshua and Millman, K. Jarrod and Mayorov,
  Nikolay and Nelson, Andrew R. J. and Jones, Eric and Kern, Robert and Larson,
  Eric and Carey, CJ and Polat, Ä°lhan and Feng, Yu and Moore, Eric W. and
  VanderPlas, Jake and Laxalde, Denis and Perktold, Josef and Cimrman, Robert
  and Henriksen, Ian and Quintero, E. A. and Harris, Charles R. and Archibald,
  Anne M. and Ribeiro, Antônio H. and Pedregosa, Fabian and {van Mulbregt},
  Paul and {SciPy 1.0 Contributors}}, ``{SciPy}: Open source scientific tools
  for python,'' 2020. [Online]. Available: \url{https://www.scipy.org/}
\BIBentrySTDinterwordspacing

\bibitem{saad1990sparskit}
Y.~Saad, ``Sparskit: A basic tool kit for sparse matrix computations,'' Tech.
  Rep., 1990.

\bibitem{li2013smat}
J.~Li, G.~Tan, M.~Chen, and N.~Sun, ``Smat: An input adaptive auto-tuner for
  sparse matrix-vector multiplication,'' in \emph{Proceedings of the 34th ACM
  SIGPLAN conference on Programming language design and implementation}, 2013,
  pp. 117--126.

\bibitem{blackford2002updated}
L.~S. Blackford, A.~Petitet, R.~Pozo, K.~Remington, R.~C. Whaley, J.~Demmel,
  J.~Dongarra, I.~Duff, S.~Hammarling, G.~Henry \emph{et~al.}, ``An updated set
  of basic linear algebra subprograms (blas),'' \emph{ACM Transactions on
  Mathematical Software}, vol.~28, no.~2, pp. 135--151, 2002.

\bibitem{10.1145/3581784.3607089}
\BIBentryALTinterwordspacing
S.~Atchley, C.~Zimmer, J.~Lange, D.~Bernholdt, V.~Melesse~Vergara, T.~Beck,
  M.~Brim, R.~Budiardja, S.~Chandrasekaran, M.~Eisenbach, T.~Evans, M.~Ezell,
  N.~Frontiere, A.~Georgiadou, J.~Glenski, P.~Grete, S.~Hamilton, J.~Holmen,
  A.~Huebl, D.~Jacobson, W.~Joubert, K.~Mcmahon, E.~Merzari, S.~Moore,
  A.~Myers, S.~Nichols, S.~Oral, T.~Papatheodore, D.~Perez, D.~M. Rogers,
  E.~Schneider, J.-L. Vay, and P.~K. Yeung, ``Frontier: Exploring exascale,''
  in \emph{Proceedings of the International Conference for High Performance
  Computing, Networking, Storage and Analysis}, ser. SC '23.\hskip 1em plus
  0.5em minus 0.4em\relax New York, NY, USA: Association for Computing
  Machinery, 2023. [Online]. Available:
  \url{https://doi.org/10.1145/3581784.3607089}
\BIBentrySTDinterwordspacing

\bibitem{cuda-quantum}
T.~C.~Q. development team, ``{CUDA Quantum},''
  \url{https://github.com/NVIDIA/cuda-quantum}, 2023, software.

\bibitem{cirq_developers_2023_10247207}
C.~Developers, ``{Cirq},'' \url{https://doi.org/10.5281/zenodo.10247207}, Dec.
  2023, software.

\bibitem{suhandli2023}
I.-S. Suh and A.~Li, ``Simulating quantum systems with {\rm nwq-sim} on {\rm
  hpc},'' in \emph{SC'23: Proceedings of the International Conference for High
  Performance Computing, Networking, Storage and Analysis}.\hskip 1em plus
  0.5em minus 0.4em\relax IEEE, 2023, p. rpost195.

\bibitem{quantum_ai_team_and_collaborators_2020_4023103}
\BIBentryALTinterwordspacing
Q.~A. team and collaborators, ``qsim,'' Sep. 2020. [Online]. Available:
  \url{https://doi.org/10.5281/zenodo.4023103}
\BIBentrySTDinterwordspacing

\bibitem{Gray2018}
\BIBentryALTinterwordspacing
J.~Gray, ``quimb: A python package for quantum information and many-body
  calculations,'' \emph{Journal of Open Source Software}, vol.~3, no.~29, p.
  819, 2018. [Online]. Available: \url{https://doi.org/10.21105/joss.00819}
\BIBentrySTDinterwordspacing

\bibitem{Suzuki_2021}
\BIBentryALTinterwordspacing
Y.~Suzuki, Y.~Kawase, Y.~Masumura, Y.~Hiraga, M.~Nakadai, J.~Chen, K.~M.
  Nakanishi, K.~Mitarai, R.~Imai, S.~Tamiya, T.~Yamamoto, T.~Yan, T.~Kawakubo,
  Y.~O. Nakagawa, Y.~Ibe, Y.~Zhang, H.~Yamashita, H.~Yoshimura, A.~Hayashi, and
  K.~Fujii, ``Qulacs: a fast and versatile quantum circuit simulator for
  research purpose,'' \emph{Quantum}, vol.~5, p. 559, Oct. 2021. [Online].
  Available: \url{http://dx.doi.org/10.22331/q-2021-10-06-559}
\BIBentrySTDinterwordspacing

\bibitem{Gidney_2021}
\BIBentryALTinterwordspacing
C.~Gidney, ``Stim: a fast stabilizer circuit simulator,'' \emph{Quantum},
  vol.~5, p. 497, Jul. 2021. [Online]. Available:
  \url{http://dx.doi.org/10.22331/q-2021-07-06-497}
\BIBentrySTDinterwordspacing

\bibitem{bayraktar2023cuquantum}
H.~Bayraktar, A.~Charara, D.~Clark, S.~Cohen, T.~Costa, Y.-L.~L. Fang, Y.~Gao,
  J.~Guan, J.~Gunnels, A.~Haidar, A.~Hehn, M.~Hohnerbach, M.~Jones, T.~Lubowe,
  D.~Lyakh, S.~Morino, P.~Springer, S.~Stanwyck, I.~Terentyev, S.~Varadhan,
  J.~Wong, and T.~Yamaguchi, ``cuquantum sdk: A high-performance library for
  accelerating quantum science,'' 2023.

\bibitem{quantuloop}
\BIBentryALTinterwordspacing
``{Quantuloop},'' \url{https://simulator.quantuloop.com}, Quantuloop, software.
  [Online]. Available: \url{https://simulator.quantuloop.com}
\BIBentrySTDinterwordspacing

\bibitem{jaques2021leveraging}
S.~Jaques and T.~Häner, ``Leveraging state sparsity for more efficient quantum
  simulations,'' 2021.

\bibitem{kaleidoscope}
T.~Dao, N.~Sohoni, A.~Gu, M.~Eichhorn, A.~Blonder, M.~Leszczynski, A.~Rudra,
  and C.~Ré, ``Kaleidoscope: An efficient, learnable representation for all
  structured linear maps,'' 12 2020.

\end{thebibliography}
\bibliographystyle{IEEEtran}



\end{document}